\documentclass[sigconf]{acmart}

\usepackage{bm}
\usepackage{tabulary}
\usepackage{multirow}
\usepackage{threeparttable}
\usepackage{caption}
\usepackage{verbatimbox}
\usepackage{subfigure}
\usepackage{graphicx}
\usepackage{url}
\AtBeginDocument{%
  \providecommand\BibTeX{{%
    \normalfont B\kern-0.5em{\scshape i\kern-0.25em b}\kern-0.8em\TeX}}}

\begin{document}

\title{Spatial Autoregressive Coding for Graph Neural Recommendation}
% \\Graph-based Recommender Systems}
\author{Jiayi Zheng}
\authornote{Both authors contributed equally to this research.}
\email{zhengjiayi980@126.com}
\affiliation{%
  \institution{Peking University}
  \city{Beijing}
  \country{China}
}
\author{Ling Yang}
\authornotemark[1]
\email{yangling0818@163.com}
\affiliation{%
  \institution{Peking University}
  \city{Beijing}
  \country{China}
}
\author{Heyuan Wang}
\email{heyuanww@163.com}
\affiliation{%
  \institution{Peking University}
  \city{Beijing}
  \country{China}
}

\author{Cheng Yang}
\email{yangcheng.iron@bytedance.com}
\affiliation{%
  \institution{ByteDance Inc}
  \city{Beijing}
  \country{China}
}
\author{Yinghong Li}
\email{liyinghong@bytedance.com}
\affiliation{%
  \institution{ByteDance Inc}
  \city{Beijing}
  \country{China}
}
\author{Xiaowei Hu}
\email{xiaowei.hu@mail.nankai.edu.cn}
\affiliation{%
  \institution{ByteDance Inc}
  \city{Beijing}
  \country{China}
}

\author{Shenda Hong}
\email{hongshenda@pku.edu.cn}
\affiliation{%
  \institution{Peking University}
  \city{Beijing}
  \country{China}
}
% \affiliation{%

% \affiliation{%

% \affiliation{%

% \affiliation{%

% \affiliation{%

% \affiliation{%


\begin{abstract}
Graph embedding methods including traditional shallow models and deep Graph Neural Networks (GNNs) have led to promising applications in recommendation.
Nevertheless, shallow models especially random-walk-based algorithms fail to adequately exploit neighbor proximity in sampled subgraphs or sequences due to their optimization paradigm. GNN-based algorithms suffer from the insufficient utilization of high-order information and easily cause over-smoothing problems when stacking too much layers, which may deteriorate the recommendations of low-degree (long-tail) items, limiting the expressiveness and scalability.

In this paper, we propose a novel framework SAC, namely Spatial Autoregressive Coding, to solve the above problems in a unified way.
To adequately leverage neighbor proximity and high-order information, we design a novel spatial autoregressive paradigm.
Specifically, we first randomly mask multi-hop neighbors and embed the target node by integrating all other surrounding neighbors with an explicit multi-hop attention.
Then we reinforce the model to learn a neighbor-predictive coding for the target node by contrasting the coding and the masked neighbors' embedding, equipped with a new hard negative sampling strategy.
To learn the minimal sufficient representation for the target-to-neighbor prediction task and remove the redundancy of neighbors, we devise Neighbor Information Bottleneck by maximizing the mutual information between target predictive coding and the masked neighbors' embedding, and simultaneously constraining 
% the mutual information
those
between the coding and surrounding neighbors' embedding. Experimental results on both public recommendation datasets and a real scenario web-scale dataset Douyin-Friend-Recommendation demonstrate the superiority of SAC compared with state-of-the-art methods.
\end{abstract}








\maketitle

\section{Introduction}
\label{intro}
% mask on 12.14
% Embedding based methods are the fundamental of modern recommender systems, which vectorize entities such as users and products. Collaborative Filtering(CF) is a representative embedding based technique without explicitly using grpah structure in the earlier research of recommmender systems. The main idea of CF is using low-dimensional vectors to characterize users and items by reconstructing the similarity of them from the historical interactions. Matrix factorization (MF) \cite{koren2009matrix} constructed embedding indexed by the unique ID of users or items and performed inner product between users and items to represent their interactions. Later on, to enrich the content of the embedding, information such as users' historical actions, social relations was adopted \cite{koren2008factorization, wang2017item}. Deep learning techniques are also widely used to make up for the shortcomings of shallow models that hard to model the complex relationships of items and users \cite{he2017neural, hsieh2017collaborative,wu2016collaborative}. 

Embedding-based methods are fundamental to modern recommender systems, which vectorize entities such as users and items based on historical interactions. Collaborative Filtering (CF) is one of the most representative techniques that use low-dimensional vectors to characterize users and items by reconstructing the similarity between them. 
% There are abundant works based on CF. 
With the development of neural networks, deep learning methods are adopted in many areas \cite{oord2018representation,he2021masked,tishby2015deep,yang2022unsupervised}, in addition to abundant works based on CF, many deep learning methods have been proposed for recommendation, covering the shortage of traditional models \cite{koren2008factorization, wang2017item,he2017neural, hsieh2017collaborative,wu2016collaborative}.

Recently, it has been shown that applying graph structure to model the interactions in recommender systems can obtain great benefits 
% \cite{ahmed2013distributed,cao2015grarep,cao2016deep,grover2016node2vec,ou2016asymmetric,perozzi2014deepwalk,song2009scalable,tang2009relational,tang2015line,wang2016structural,wang2016structural,wang2017community,wei2017cross,zhou2017scalable}. 
\cite{grover2016node2vec,perozzi2014deepwalk,tang2015line,wang2016structural,wang2017community,wei2017cross}. 
The most common paradigm of graph-based recommendation is introducing representations for nodes with graph embedding techniques. 
% Traditional shallow graph embedding methods mostly utilize a random walk process to construct local context to preserving the proximity in the low-dimensional space\cite{perozzi2014deepwalk,grover2016node2vec,ribeiro2017struc2vec,tang2015line}. 
Traditional shallow models mostly utilize matrix factorization or random walk process to construct local context to preserve 
% the proximity
the adjacency similarity or structural similarity
in the low-dimensional space \cite{perozzi2014deepwalk,grover2016node2vec,ribeiro2017struc2vec,tang2015line}.
% Through the graph embedding process, various downstream recommendation tasks can be performed
Many works based on shallow graph embedding models for recommendation have achieved good results
\cite{yang2018hop,wang2018billion,zhang2019neuralPKU}. 
Recent years, 
% among those graph embedding algorithms, 
% with the development of deep learning,
many deep models for graph embedding emerged, among which Graph Neural Networks (GNNs) received 
% more and more attention
special attention
because of their superior ability in learning graph-structured data. The main idea of these GNN-based methods is to iteratively aggregate feature information from neighbors and integrate it with the current central node representation
% during the propagation process 
\cite{kipf2016semi,velivckovic2017graph,hamilton2017inductive,wu2020comprehensive,berg2017graph,ying2018graph,yang2020dpgn,pmlr-v162-yang22d}. 
% With the rapid developing of deep learning, 
Numerous GNN-based methods been proposed for recommendation, and the performances of these methods far surpass traditional algorithms \cite{wu2020graph, ying2018graph,he2020lightgcn,wu2021self}.
% \begin{figure*}[h]
% \centering
%     \includegraphics[width=8cm]{kdd2022.pdf}
% \caption{} 
% \label{fig:f1}
% \end{figure*}
\begin{figure}
\setlength{\abovecaptionskip}{0cm}
\centering
    \includegraphics[width=8cm]{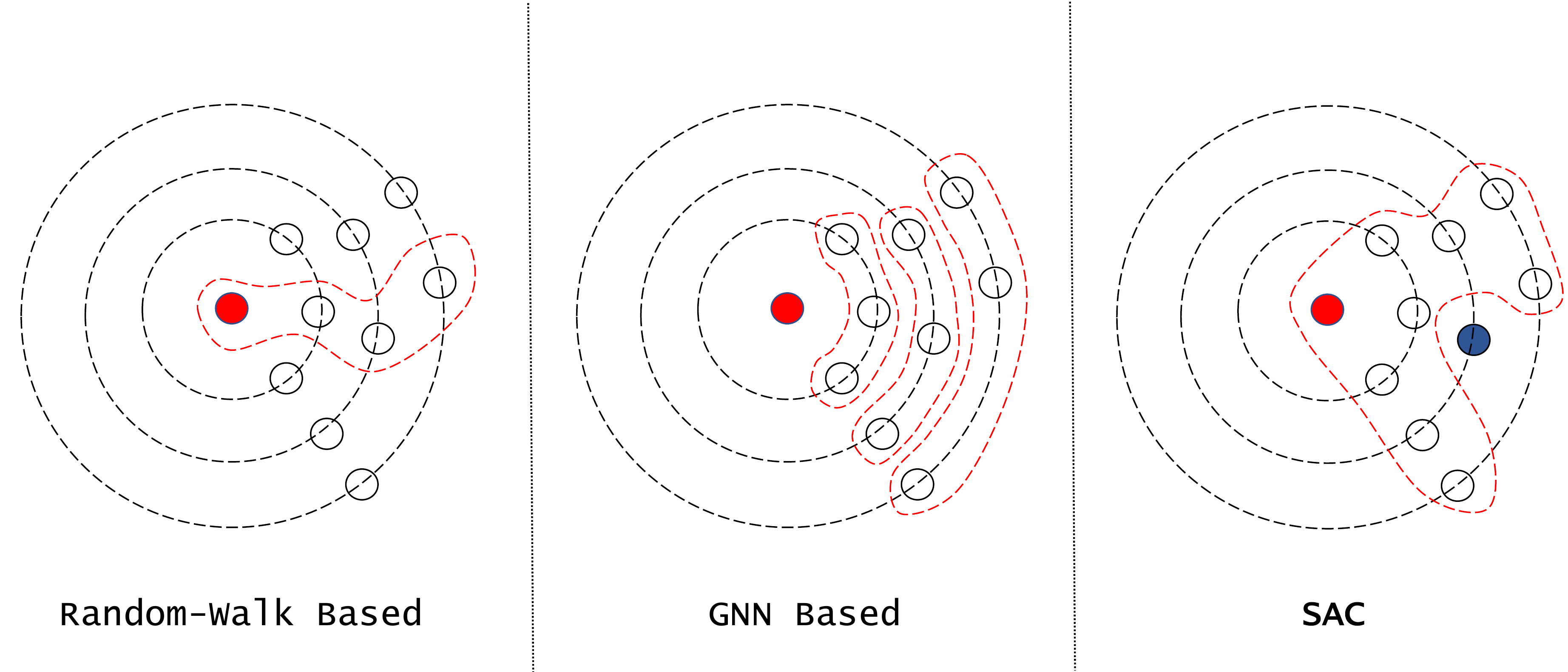}
\caption{Schematic comparisons between our proposed SAC and previous methods in graph embedding process. SAC learns the embedding of target node (in red) by predicting the masked neighbor (in blue) in a spatial autoregressive way.}
\label{fig:f1}
\end{figure}
Despite their effectiveness in recommendation,
% and other tasks, 
% we argue that both random-walk-based and GNN-based methods 
we argue that both traditional shallow and deep models
are not sufficient to learn good enough representations especially when scaling them up to large scale graph with sparse connections. We go deep into their characteristics and point out their limitations as the following:
\begin{itemize}
    % \item Random walk sampling of different lengths is involved with varying scales. Nevertheless, downstream task performance of embeddings has been shown to be insensitive to lengths since abundant neighbor context is inadequately utilized, which limits the expressiveness of embeddings.
    % \item Random walk sampling has limited capacity to model the nodes' context information, which limits the expressiveness of embedding. Moreover, they are prone to be affected by sampling bias and thus need to design some complicated strategies to deal with this problem.
    \item Shallow models mainly rely on matrix decomposition or random walk. Matrix decomposition usually has high time complexity and cannot be extended to large-scale graphs. Most random-walk-based shallow models have limited capacity to model the neighbor proximity because only a small part of neighbors are sampled as described in Figure \ref{fig:f1}, and they are prone to be affected by sampling bias.
    
    % \item GNN-based methods usually conduct feature propagation across hops through stacked GNN layers. They are unable to simultaneously process multi-hop feature in an explicit way. Besides, the layers of them can not be too deep since it may cause over-smoothing problem, which limits its integration of high-order information for target node representation. 
    
    % \item Most of the GNN-based methods for recommendation failed to work when scaling up to web-scale graphs because of the impossibility for loading the whole graph Laplacian matrix into memory. 
    \item Some deep models only exploit deep encoders 
    % and still have limited capacities for modeling the context 
    and model the context in a similar way to random walk,  still limited in capability
    \cite{wang2016structural, cao2016deep}. While other GNN-based models conduct feature propagation across hops
    % implicitly layer by layer,
    iteratively,
    which causes over-smoothing problems when stacking too many layers, limiting the ability for integrating high-order information. 
    % The GNN paradigm 
    This limitation also amplifies the influence of high degree nodes, causing long-tail nodes cannot capture sufficient information.
    % since too many layers may cause over-smoothing problems, 
    % it limits the ability for integrating high-order information. 
    In addition, loading the whole graph Laplacian matrix into memory is impossible for large-scale graphs, at which point most GNN models \cite{ he2020lightgcn, wu2021self, kipf2016semi, velivckovic2017graph} fail to work.
    % And their two-tower architecture for encoding user-item pairs leads to heavy computational load, which also make the model not applicable in practice when scaling up to large-scale recommender systems.
    % Inadequate utilization of high-order information makes existing methods perform not so well for large-scale recommender systems where there are many low-degree nodes desiring high-order proximity. 
\end{itemize}

\subparagraph{\textbf{Present work.}}%To this end,%
We propose SAC, Spatial Autoregressive Coding, a novel and effective framework to address aforementioned problems in a unified way, as illustrated in Figure \ref{fig:f1}. 
% We aim to learn neighbor proximity preserved embeddings, while in large-scale recommender systems, most nodes in the network have sparse connections (i.e. long-tail phenomenon) and thus such nodes need to aggregate high-order information to learn enough expressiveness.
% With these in mind, 
% we devise a new spatial autoregressive (target-to-neighbor) embedding paradigm,
% utilizing abundant neighbor proximity and multi-hop context with a single network.
Specifically, 
% for adequately exploiting neighbor proximity,
we first use node-wise sampling to sample multi-hop neighbors for 
% each 
target node. Then we perform random masking on neighbors at each hop and flatten all other surrounding neighbors along with the target node to a single Transformer-based encoder, which directly integrates the multi-hop neighbors context into target node representation while alleviating over-smoothing caused by layer-by-layer aggregation. After getting the update target, we apply an autoregressive model to reinforce the target node to approximate the masked neighbors
% by straightly minimizing contrastive loss. 
in a contrastive manner.
% straightly optimizing a spatial neighbor-predictive coding.
The whole masking-and-predicting process induces the latent space to capture information that is maximally useful to predict neighbor nodes and preserve structural proximity 
and high-order information
in a unified way. 
% By approximating multi-hop neighbors, the target node's representation can better utilize high-order information, which helps enhance context information for low degree nodes.
To improve the robustness of the model, we equip the contrastive loss with a novel hard negative sampling strategy, which helps to better understand the boundary between positive and negative samples.
It is noted that we also devise Neighbor Information Bottleneck (NIB) to remove the information redundancy in neighbor aggregation process and thus learn a minimal sufficient representation for prediction task. NIB is realized by maximizing the mutual information between target coding and the masked neighbors and applying constraint on the mutual information between target coding and surrounding neighbors. Empirical experiments are conducted on datasets of multiple scales to prove the effectiveness and scalability of SAC.

To summarize, our paper makes the following contributions:
\begin{itemize}
    \item To the best of our knowledge, we firstly propose Spatial Autoregressive Coding (SAC) for recommendation, to learn a neighbor-predictive coding for target node 
    by adequately utilizing neighbor proximity
    in a spatial autoregressive paradigm, which effectively maximizes
    the mutual information of target-neighbor pairs in latent space. 
    % and adequately utilizes high-order information.
    \item We are the first to propose Multi-Hop Neighbor Modeling to explicitly integrate complete neighbor context into target node with long-range attention, and approximate multi-hop neighbors simultaneously, which better models high-order information and alleviates the long-tail problems.
    % alleviates the problems of sparse behaviors of long-tail users in recommendation. 
    \item We propose a negative sampling strategy for graph-based contrastive learning, using random walks and distance-based similarity metrics to generate hard negatives, helping distinguish between positive and negative samples better.
    \item To remove the information redundancy in neighbor propagation process, we propose Neighbor Information Bottleneck to learn the \textit{minimal} sufficient representation for target node. It discourages the representation from acquiring additional information from the surrounding neighbors that is irrelevant for predicting the neighbor (\textit{minimal}).
    % \item Our algorithm just utilizes single encoder to realize the whole network embedding framework, which is computationally efficient and makes it scalable to large-scale recommender systems compared to previous two-tower based methods.
    \item We conduct experiments on three public large-scale datasets while most of the previous methods choose smaller ones. We also verify SAC on a web-scale dataset Douyin-Friend-Recommendation to evaluate the model in a real scenario. Experimental results show SAC outperforms previous methods by a significant margin. We also conduct ablation studies to demonstrate the effectiveness of SAC. 
\end{itemize}

\begin{figure*}[h]
\centering
    \includegraphics[width=13cm]{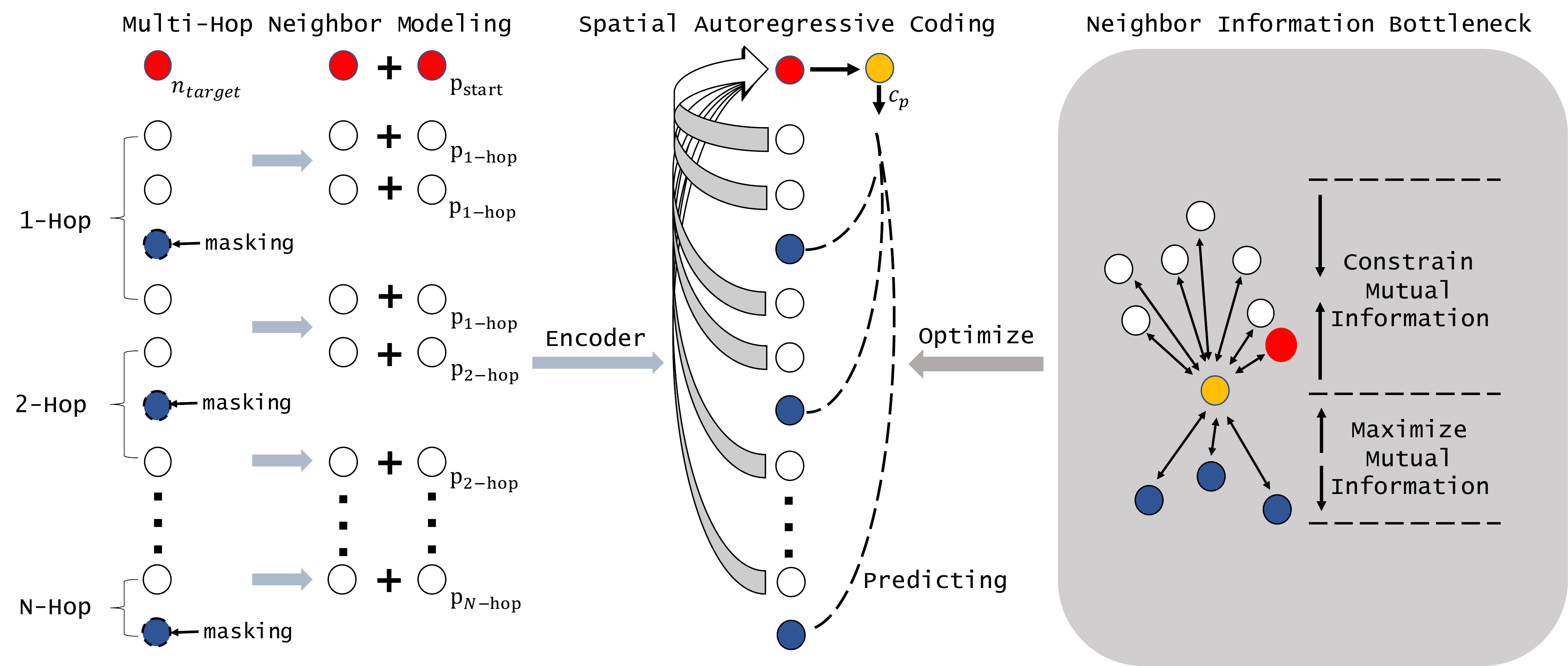}
\caption{The illustration of our SAC. The Multi-Hop Masking is applied in neighbors centered around the target node. After the encoding process, SAC produces the neighbor-predictive coding for the target and takes the masked neighbors as the positive samples. Finally, SAC is simultaneously optimized by the contrastive loss and Neighbor Information Bottleneck.} 
\label{fig:f2}
\end{figure*}

\section{Related Work}
We briefly review existing studies on garph neural recommendation related to our work: traditional CF-based methods, graph embedding methods including shallow and deep models. 
% Besides, we simply introduce autoregressive models.
\\
\subparagraph{\textbf{Traditional CF-based methods for recommendation.}} Collaborative Filtering (CF) is a representative embedding based technique.
% without explicitly using grpah structure
% in the earlier research of recommmender systems. 
Matrix factorization (MF) \cite{koren2009matrix} constructed embedding indexed by the unique ID of users or items and performed inner product between them to represent their interactions. 
Later on, to enrich the content of the embedding, information such as users' historical actions, social relations was adopted \cite{koren2008factorization, wang2017item}. 
Deep learning techniques are also widely used
to make up for the shortcomings of shallow models
% that hard to model the complex relationships of items and users
\cite{he2017neural, hsieh2017collaborative,wu2016collaborative}. Neural Factorization Machine \cite{he2017neural} combines second-order linear features extracted by FM \cite{rendle2010factorization} with the high-order nonlinear features extracted by neural network. 
% In order to tackle the problem that traditional CF can only model the similarity of user-item pairs while ignoring the similarity of user-user and item-item pairs, 
\cite{hsieh2017collaborative} exploit metric learning to enhance the CF model. \cite{wu2016collaborative} further using auto-encoders to achieve a better result for top-n recommendation.
\\
% \subparagraph{\textbf{Shallow graph embedding methods.}}
\subparagraph{\textbf{Graph embedding methods for recommendation.}}
Most graph embedding methods can be divided into two categories: shallow models and deep models.
Traditional shallow models mainly exploit matrix factorization or random walk.
% and \cite{qiu2018network} points out that random walk methods can be unified under the framework of matrix factorization theory.
Random-walk-based methods utilize random walk process to sample nodes and some of them rely on the Skip-Gram language model \cite{mikolov2013distributed}, which aim to capture the pointwise similarity \cite{levy2014neural}.
% Classic works such as DeepWalk \cite{perozzi2014deepwalk}, Node2vec \cite{grover2016node2vec}, LINE \cite{tang2015line} all use a random walk process to learn a structure-preserved representations.
% DeepWalk \cite{perozzi2014deepwalk} uses local information obtained from truncated random walks to learn latent representations. 
% by treating walks as the equivalent of sentences.
% Node2vec \cite{grover2016node2vec} proposes an efficient $2^{nd}$ random walk algorithm.
% proposes an efficient scalable algorithm for feature learning in networks that efficiently optimizes a novel network-aware, neighborhood preserving objective using SGD. 
% Struc2vec \cite{ribeiro2017struc2vec} uses a hierarchy to measure node similarity at different scales.
% , and constructs a multi-layer graph to encode structural similarities and generate structural context for nodes.
% LINE \cite{tang2015line} optimizes a carefully designed objective that preserves both local and global structures.
% APP \cite{zhou2017scalable} proposes an asymmetric proximity preserving graph embedding method via random walk with restart. 
There are abundant random-walk-based works for recommendation.
HOP-Rec \cite{yang2018hop} combines the embedding-based MF methods with random walks to enrich user's context. \cite{wang2018billion} also trains the embedding by combining random walks and Skip-Gram model \cite{mikolov2013distributed} to generate nodes' rich context for click through rate (CTR) prediction. GEPS \cite{zhang2019neuralPKU} exploits DeepWalk  \cite{perozzi2014deepwalk} and Node2Vec \cite{grover2016node2vec} to integrate neighbors' information. 
% \subparagraph{\textbf{Deep graph embedding methods.}}
In recent years, many works introduce deep models into graph embedding \cite{yang2022omni}. For example, auto-encoder is introduced with random walk for recommendation \cite{wang2016structural, cao2016deep}. From the perspective of user's historical behavior and time series modeling \cite{yang2022unsupervised-arxiv}, Transformer is also applied for sequential recommendation \cite{xiao2021uprec, kang2018self}. 
Among those, GNNs achieve great success.
GCN \cite{kipf2016semi} approximates the first-order eigen-decomposition of the graph Laplacian to iteratively aggregate information from neighbors. 
Graph Attention Network (GAT) \cite{velivckovic2017graph} further proposed attention-based neighbor aggregation.
% by calculating the correlation between the current node and neighbors.
GraphSage \cite{hamilton2017inductive} samples a fixed size of neighbors for each node, aims at learning the aggregating function.
% which can be deployed on large-scale systems.
% ClusterGCN \cite{chiang2019cluster} proposes a novel GCN training algorithm by exploiting the graph clustering structure.
% GraphSAINT \cite{zeng2019graphsaint} introduces a graph sampling based inductive learning method.
% that improves training efficiency and accuracy in a fundamentally different way.
GNN-based graph embedding methods has made great improvements in many tasks, leading to a surge of works that applying GNNs to recommender systems.
PinSage \cite{ying2018graph} combines efficient random walks and graph convolutions to generate embeddings of nodes (i.e. items)
that incorporate both graph structure as well as node feature information, it also proposed a hard negative sampling method. LightGCN \cite{he2020lightgcn} learns user and item embeddings by linearly propagating them 
on interaction graph, 
and uses the weighted sum of the embeddings learned at all layers as the final embedding. SGL \cite{wu2021self} proposed a self-supervised framework, using data augmentation to produce multi-views for nodes and their context, then GCN is adopted to generate the embedding.
GIN \cite{li2019graph} adopts multi-layered graph diffusion to enrich user behaviors for solving the behavior sparsity problem. 

However, these 
works have very limited scalability, and
cannot fully model the neighbor 
proximity while effectively integrating higher-order information, that's what we're trying to solve.
\section{Preliminary}
\label{section_preliminary}
We introduce the common paradigm of GNN-based recommendation methods. Let $\mathcal{U}$ and $\mathcal{I}$ be the set of users and items respectively, for predicting how likely user $u\in\mathcal{U}$ would adopt item $i\in\mathcal{I}$, we produce representation $\mathbf{n}_{u}$ and $\mathbf{n}_{i}$  by aggregating neighbors iteratively
with an $L$-layers GNN
on bipartite graph $\mathcal{G}=(\mathcal{V}, \mathcal{E})$, in which $\mathcal{V}=\mathcal{U}\cup\mathcal{I}$ and $\mathcal{E}$ contains all the interactions:
\begin{equation}
    \mathbf{n}_{u}^{(l)}=COMBINE(\mathbf{n}_{u}^{(l-1)}, \, AGG(\{\mathbf{n}_{j}^{(l-1)}|j\in\mathcal{N}_u\})),
\end{equation}
where $\mathcal{N}_u$ denotes the neighbors of $u$, $COMBINE$ and $AGG$ denotes the combing function and aggregating function respectively. A readout function is optional to generate the final $\mathbf{n}_u$ by combing $\mathbf{n}_u^{(j)}, j=1,2,...,L$. Commonly, we predict the preference score of $u$ on item $i$ by computing the inner product of their representations, i.e. $s_{ui}=\mathbf{n}_u^T\mathbf{n}_i$. In different scenarios, items are different types of things, such as friends, goods, etc. To better distinguish between different node types in the graph, in the following, we denote the target user as $n_{target}$, the multi-hop neighbors of the target user as $n_1, n_2, ..., n_k$. The neighbors are composed of both similar users and items. Unless otherwise specified, the bold character represents the embedding of the corresponding node symbolized by that character.

\section{Spatial Autoregressive Coding}
In this section, we introduce our proposed algorithm in detail. Subsection \ref{SAP} introduces vanilla Spatial Autoregressive Coding (SAC) including new Spatial Autoregressive Paradigm for graph embedding, powered by Multi-Hop Neighbor Modeling. Subsection \ref{NIB} introduces the Neighbor Information Bottleneck for enhancing the SAC and provides theoretical analyses. 
Subsection \ref{model-opt} introduces a new negative sampling method for training and give the final optimization objective for SAC.

\subsection{Spatial Autoregressive Paradigm}
\label{SAP}

% \subsubsection{\textbf{Spatial Autoregressive Paradigm.}}
% \label{subsub-spatial-autoregressive-paradigm}
\subsubsection{\textbf{Neighbor Predictive Learning.}}
\label{NCE_describe}The main intuition of our Spatial Autoregressive Paradigm is that embedding all the shared neighbor proximity into the node representations among the subgraph. 
% while discarding instance-level noise that is useless.
We convert the previous graph embedding task into a neighbor prediction task since in high-dimensional graph-structured data modeling, using neighbor prediction is able to adequately exploit the local smoothness and neighbour contexts of the target node. When predicting higher order of neighbor, the amount of shared information becomes much lower, and the model needs to capture high order proximity.
For example, in the recommender systems, 
% we usually need to find underlying possible interests (items or persons).
the graph structure is usually user-item bipartite graph with extremely low sparsity.
% The edges in the graph usually denoted the observed interactions.
Therefore, capturing high order proximity is also non-trivial to improve the performance of recommendation, which presents a way to find underlying possible interests for users.
Figure \ref{fig:f2} shows the architecture of spatial autoregressive coding algorithms.
% Considering a sampled $k$-rooted subgraph $(\mathcal{G},n_1,\cdots,n_k)$, 
% Let $\mathcal{U}$ and $\mathcal{I}$ denote the set of users and items respectively, 
We select a target node $n_{target}$ and sample N-hop neighbors of $n_{target}$ to form a subgraph, denoted by $(\mathcal{G},n_1,\cdots,n_k)$. Then we integrate target node $n_{target}$ with all other surrounding neighbors to predict a randomly masked neighbor node $n_i$. Thus the problem is to estimate the likelihood:
\begin{equation}
    Pr(\{\mathbf{n}_{i}|\mathbf{n}_{target},\ \mathbf{n}_1,\ \cdots,\ \mathbf{n}_{i-1},\ \mathbf{n}_{i+1},\ \cdots,\ \mathbf{n}_k\})
\end{equation}
where $\mathbf{n}$ represents the trainable node embedding
% \footnote{In the following, unless otherwise specified, the bold character represents the embedding of the corresponding node symbolized by that character.}
and $i$ is the index of the masked neighbor. 
The spatial autoregressive process expects the model to learn a neighbor-predictive coding to maximize the likelihood function. To achieve this, we aim to learn a neighbor predictive coding to predict the masked neighbor:
\begin{equation}
\begin{split}
    \label{eq-autoregressive}
    \mathbf{c}_{p}=&f_{\theta}(\{\mathbf{n}_{target}+\mathbf{p}_{target},\ \mathbf{n}_1+\mathbf{p}_1,\ \cdots,\\
    &\mathbf{n}_{i-1}+\mathbf{p}_{i-1},\ \mathbf{n}_{i+1}+\mathbf{p}_{i+1},\ \cdots,\ \mathbf{n}_k+\mathbf{p}_{k}\}),
\end{split}
\end{equation}
where $\mathbf{p}$ represents the hop-indexed position embedding and $\mathbf{c}_{p}$ is the predictive coding that aggregates the surrounding neighbors with target node. $f_{\theta}$ is the normal Transformer encoder.
% empowered by multi-head self-attention mechanism \cite{vaswani2017attention}:
% \begin{align}
%     \text{MultiHead}(Q,K,V) &= \text{Concat}(head1,head2,\cdots,head_h)W^O,
% \end{align}
% where $Q,K,V\in \mathop{\mathbb{R}}^{k\times d}$ are input node embedding matrices, $d$ is the embedding dimension, and $h$ is the number of heads. Each head is defined as the following:
% \begin{align}
%     \begin{split}
%     head_i&=Attention(QW_i^Q,KW_i^K,VW_i^V)\\
%     &=\text{softmax}\left (\frac{QW_i^Q(KW_i^K)^T}{\sqrt{d_k}}\right)\cdot VW_i^V,
% \end{split}
% \end{align}
% where $W_i^Q,W_i^K,W_i^V$ are linear projections.
The process enables target node to capture attentional context based on local proximity that is helpful to predict the masked neighbor.
To obtain the compact predictive $\mathbf{c}_p$, our model
% By producing $\mathbf{c}_p$, we aimed at predicting the masked neighbor. In other words, we want $\mathbf{c}_p$ to distinguish the masked node from all other nodes. Motivated by existing works, we follow SimCLR\cite{chen2020simple} and optimize contrasive loss.
% A good representation of $\mathbf{c}_p$ 
needs to meet the following two characteristics: 1. predicting the representation of masked neighbor node; 2. distinguishing the representation from other noisy nodes. Thus in a subgraph batch, for each target node, we treat $\mathbf{c}_p$ and the masked neighbor node ${n}_i $ in the same subgraph as a positive pair, and treat nodes beyong the subgraph as negative samples of $\mathbf{c}_p$, denoted by $\mathcal{B}^-$, and let $\mathcal{V}^{'}=\mathcal{B}^- \cup \{{n}_i\}$.
% denoted by $\mathcal{B}^-$. 
% Different from previous approaches with a two-tower training architecture, 
Thus our SAC uses a target-to-neighbor learning objective by adopting contrastive loss InfoNCE \cite{gutmann2010noise} to maximize the agreement of $\mathbf{c}_p$ and $\mathbf{n}_i$ while minimizing that of negative pairs:

{
\setlength{\abovedisplayskip}{-0.15cm}   
\begin{align}
         \label{eq-info-nce}
         \mathcal{L}_{Vanilla-SAC} = {-\text{log}\frac{\text{exp}(\mathbf{c}_p^T\mathbf{n}_i/\tau)}
         {\sum_{n'\in\mathcal{V}^{'}}{\text{exp}(\mathbf{c}_p^T\mathbf{n}'/\tau)}}}
\end{align}}
where $\tau$ is the temparature parameter. Such learning objective optimizes both the target and masked neighbor embeddings.

% old
% Different from previous approaches with a two-tower training architecture, our SAC applies an effective single encoder and simultaneously optimizes the target and masked neighbor embedding through the following objective: 
% \begin{align}
%     \label{eq-contrastive}
%     \mathcal{L}_{Vanilla-SCAT} = \frac{1}{B}\sum^{B}_{b=1}{(\mathbf{c}_{b,p}-\mathbf{n}_{b,i})}^2
%     % \mathcal{L}_{p} =
%     % -log({sim(\mathbf{c}_p,\mathbf{n}_i)/\tau}) +log({sim(\mathbf{c}_p,\mathbf{n}_j))/\tau},
% \end{align}
% where $B$ denotes the number of sampled subgraph in the mini-batch,  $\mathbf{n}_{b,i}$ is the $i$-th masked neighbor of the $b$-th subgraph and $\mathcal{L}_{Vanilla-SCAT}$ represents the prediction loss for each target node. It is noted that our Spatial Autoregressive Paradigm avoids the problem of bad negative samples since we only use nodes from the sampled subgraph to construct the loss function, guaranteeing a more effective and stable training procedure.

\subsubsection{\textbf{Multi-Hop Neighbor Modeling.}}
In recommender systems especially in large-scale ones, there usually exist a number of 
sparsely-connected entities, i.e long tail phenomenon.  Thus nodes of low degrees need high-order information to embedding the structural proximity. Although existing methods have made efforts to embed the multi-hop information into node representation, the way they conduct is implicit and usually in a cascade way, which would lead to 
% the loss
underutilization
of partial useful high-order information. Therefore, we propose the Multi-Hop Neighbor Modeling to enhance the multi-hop information processing. Specifically, for a same target node, we simultaneously mask the neighbors of multiple hops and add the relative position embedding to the rest neighbors according to the hop index. In the prediction procedure, we predict the masked neighbors all at once which reinforces the target node to explicitly and simultaneously acquire the multi-hop information. The procedure makes the neighbor predictive coding $\mathbf{c}_p$ reconstruct multi-hop local structure around $n_{target}$.  We reformulate the loss function (\ref{eq-info-nce}) as the following:
{
\begin{align}
    \label{eq-multihop}
    \mathcal{L}_{Vanilla-SAC} = \sum^N_{h=1}{-\text{log}\frac{\text{exp}(\mathbf{c}_p^T\mathbf{n}_{i, h}/\tau)}
         {\sum_{n'\in\mathcal{V}^{'}}{\text{exp}(\mathbf{c}_p^T\mathbf{n}'/\tau)}}},
    % \mathcal{L}_{p} = \sum^{N}_{hop=1}
    % -log({sim(\mathbf{c}_p,\mathbf{n}_{i,hop})/\tau}) +log({sim(\mathbf{c}_p,\mathbf{n}_j))/\tau},
\end{align}
}
where $\mathbf{n}_{i,h}$ denotes the $i$-th neighbour which is masked at hop $h$, there are a total of N-hop neighbors. The loss function $\mathcal{L}_{Vanilla-SCAT}$ takes the multi-hop optimization into account and explicitly utilizes high-order information.

% There are some significant difference between our SAC and UP-Rec:
% \begin{itemize}
%     \item We aggregate both the historical interactions of user-item and similar users(since we build the subgraph by sampling multi-hop neighbors), while UP-Rec only use the first one.
%     \item We directly generate a neighbor predictive coding for target node, which can be directly used for representing target node in downstream tasks. While UP-Rec is a pre-train framework, which do not directly model the historical interactions and need to be tuned for downstream tasks.
%     \item Multi-hop neighbor modeling means we try to model the virtual interactions between user and item while UP-Rec only use the real interactions.
% \end{itemize}

\subsection{Neighbor Information Bottleneck}
\label{NIB}
Although vanilla SAC performs well, it deteriorates the performance when encountering complex graphs which consists of diverse neighbors and complicated local topology structures. Therefore, when predicting the different masked neighbors, we should only extract the minimal sufficient information and filter the irrelevant information from the surrounding neighbors.
% We note that predicting nodes in high-dimensional graphs with unimodal losses is not very efficient, and powerful conditional generative models which need to reconstruct complete structural proximity in the graph are usually required. But these models are computationally intense, and waste capacity at modeling the complex topology. In aforementioned prediction scenario, there may exist many different surrounding neighbors while few shared information among them would help the target node to predict the masked neighbor.
To achieve this goal,
% and thus enhance vanilla SAC,
we propose Neighbor Information Bottleneck (NIB), an information-theoretic principle inherited from Information Bottleneck \cite{tishby2015deep}. NIB is designed for learning informative and predictive coding for node representation in graphs. Specifically, after we obtain the neighbor-predictive context coding $\mathbf{c}_p$, NIB maximizes the mutual information between the $\mathbf{c}_p$ and the masked neighbor $\mathbf{n}_i$, and simultaneously constrain the mutual information between the $\mathbf{c}_p$ and the input nodes (including target node and surrounding neighbours), $\mathbf{x}_{in}=\{\mathbf{n}_{target},\mathbf{n}_1,\cdots,\mathbf{n}_{i-1},\mathbf{n}_{i+1},\cdots,\mathbf{n}_k\}$. Based on this formulation, the objective can be summarized as:

{
\setlength{\abovedisplayskip}{-0.2cm}   
\setlength{\belowdisplayskip}{0cm}  
\begin{align}
\label{bottleneck}
    \mathop{max}\limits_{\theta} \quad I(\mathbf{c}_p, \mathbf{n}_i;\theta)-\beta I(\mathbf{c}_p,\mathbf{x}_{in};\theta)
\end{align}}
where $\beta$ is the Lagrange multiplier attached to the constrained meaningful
information. $\theta$ is the parameters of the Transformer encoder and we will remove it for simplicity. The first term in the formula expects the $\mathbf{c}_p$ to contain more informative information for predicting $\mathbf{n}_i$, and the second term wants $\mathbf{c}_p$ to remove the useless information from $\mathbf{x}_{in}$.

Objective (\ref{bottleneck}) contains two terms about estimating mutual information. However, the mutual information is hard to estimate and we should transform it to another way. Take $I(\mathbf{c}_p, \mathbf{n}_i)$ for example:
\begin{align}
\label{eq-mutual}
\begin{split}
    I(\mathbf{c}_p, \mathbf{n}_i)&=\int_{c\sim \mathbf{c}_p} \int_{n\sim \mathbf{n}_i}  p(n,c)\,\text{log}\frac{p(n,c)}{p(n)p(c)} \,dn\,dc\,\\
    &=\int_{c\sim \mathbf{c}_p} \int_{n\sim \mathbf{n}_i}  p(n,c)\,\text{log}\frac{p(n/c)}{p(n)}\,dn\,dc\,\\
    &=KL(p(c,n)||p(c)p(n))\\
    &\propto JS(p(c,n)||p(c)p(n)).
\end{split}
\end{align}
Note that the distributions of the nodes are hard to estimate and thus we refer to \cite{nowozin2016f,tishby2000information,tishby2015deep} for an alternative way to transform the objective (\ref{bottleneck}) and derive the NIB loss $\mathcal{L}_{NIB}$ as following:
% {
% \setlength{\abovedisplayskip}{0.1cm}   
% \setlength{\belowdisplayskip}{0.1cm} 
% \begin{equation}
% \label{NIB-loss}
% \begin{split}
%     \mathcal{L}_{NIB}=&\sum_{i=1}^{N}-\text{log}(\sigma(\mathbf{c}_{:,p}^TW_1\mathbf{n}_{:,i,h}))+\\
%     &\beta\sum_{j=1}^k\text{log}(\sigma(\mathbf{c}_{:,p}^TW_2\mathbf{n}_{:,j})),
% \end{split}
% \end{equation}
% }
{
\setlength{\abovedisplayskip}{0.1cm}   
\setlength{\belowdisplayskip}{0.1cm} 
\begin{equation}
\label{NIB-loss}
\begin{split}
    \mathcal{L}_{NIB}=&\sum_{h=1}^{N}-\text{log}(\sigma(\mathbf{c}_{p}^TW_1\mathbf{n}_{i,h}))+\beta{\sum_{\mathbf{n}_j\in\mathbf{x}_{in}}}\text{log}(\sigma(\mathbf{c}_{p}^TW_2\mathbf{n}_{j})),
\end{split}
\end{equation}
}
where $N$ is the number total hops and $k$ is the number of nodes (excluding masked neighbor) in subgraph, $\sigma$ denotes the non-linear activation function, $W_1$ and $W_2$ is linear transformation matrices for bilinear fusion of $\mathbf{c}_p$ and $\mathbf{n}_{i,hop}$. NIB focuses on optimizing the efficacy of local aggregation while Objective (\ref{eq-multihop}) aims to discriminate the representations in a global space.

\begin{figure}[]
\centering
    \includegraphics[width=7cm]{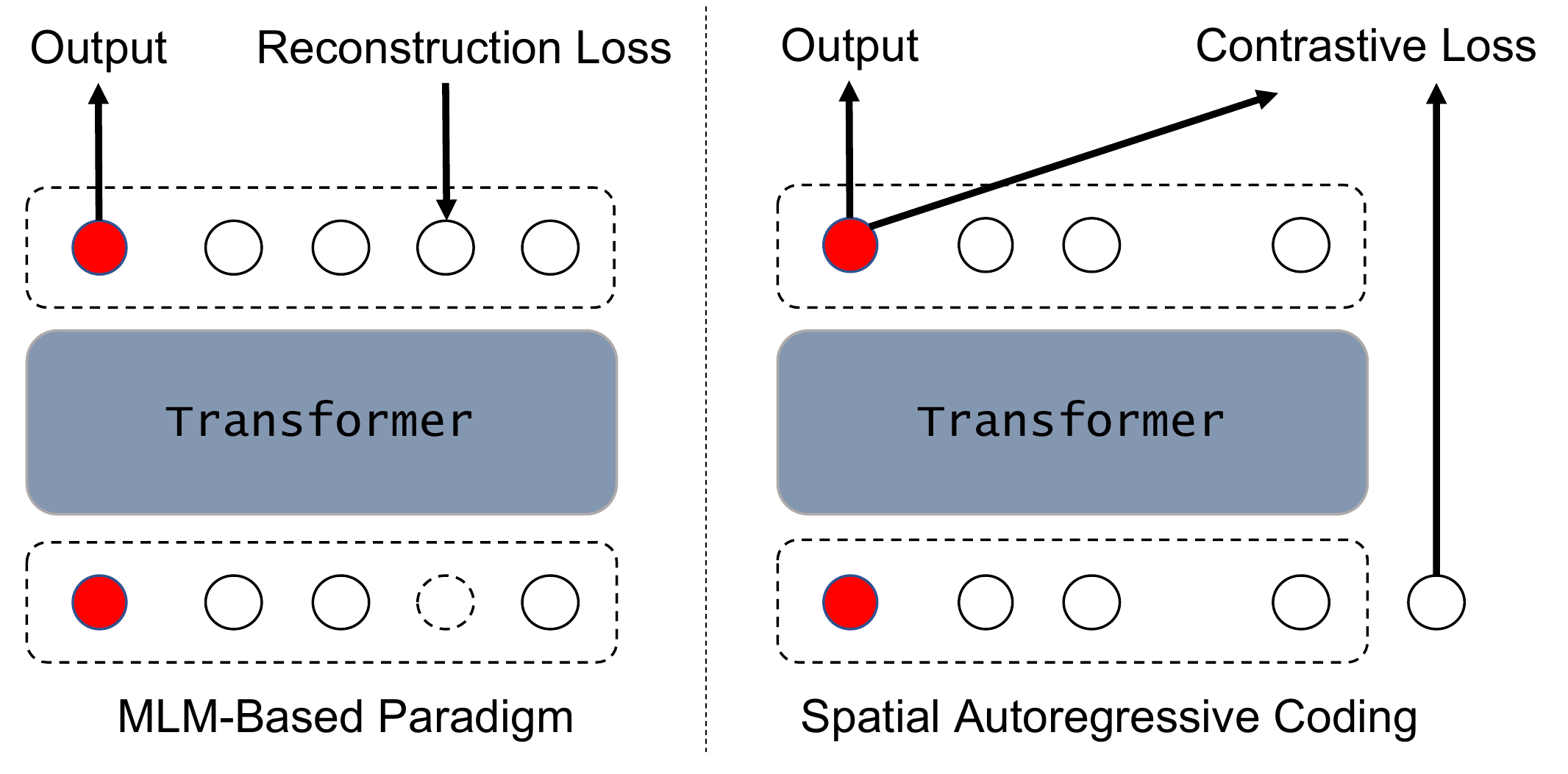}
\caption{Schematic diagram of comparisons between MLM-based paradigm and our SAC.} 
\label{fig:f3}
\end{figure}

% where $:$ denotes the batch axis. 
\subsection{Model Optimization}
\label{model-opt}
\begin{figure}
\setlength{\abovecaptionskip}{0.2cm}
\centering
    \includegraphics[width=6.5cm]{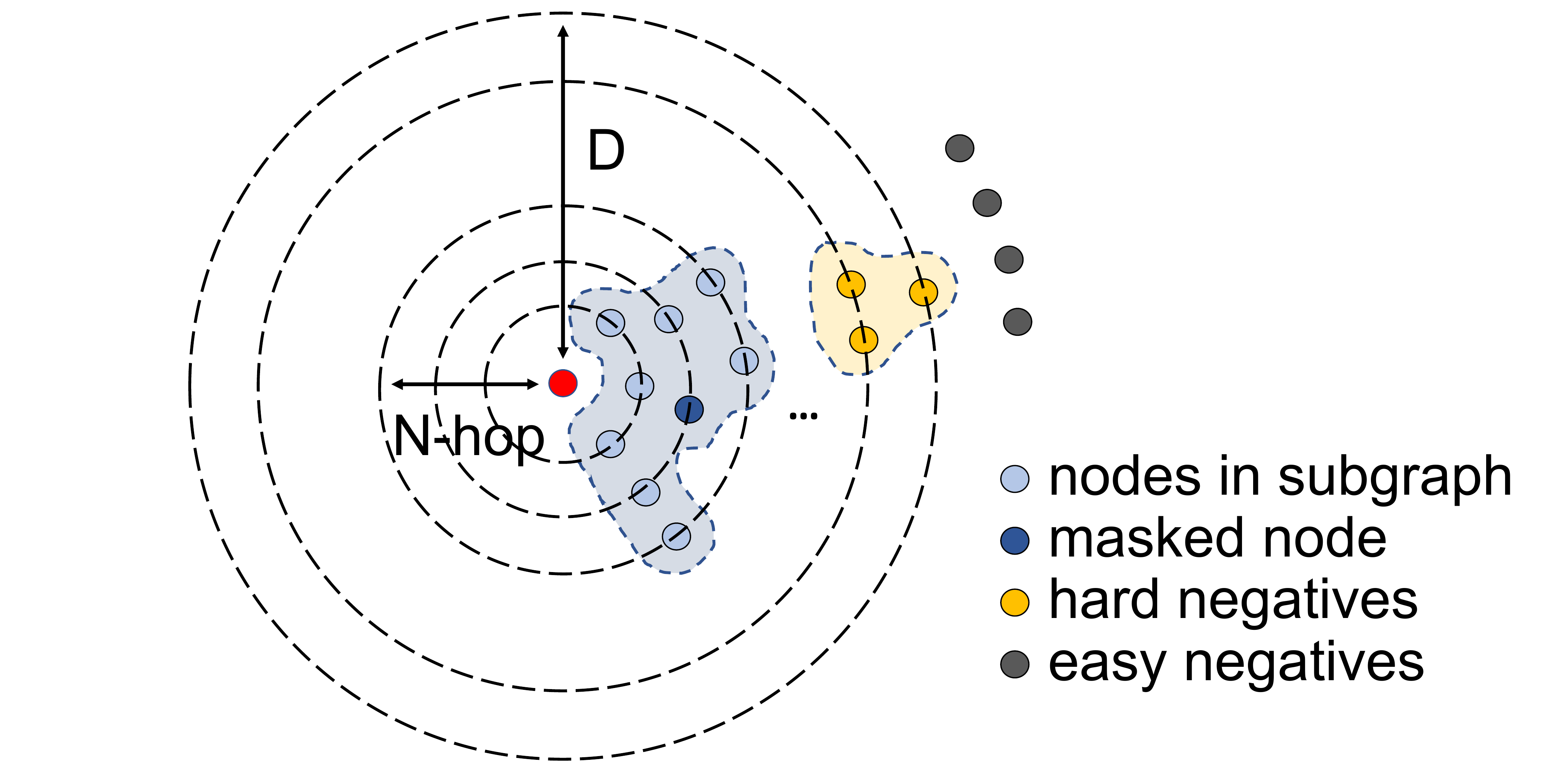}
\caption{The illustration of hard negatives sampling strategy.}
\label{fig:hn_sampling}
\end{figure}

\subparagraph{\textbf{Negative Sampling Strategy.}}  The summation item in loss function (\ref{eq-multihop}) need to be estimated by designing a negative sampling strategy. Instead of simply sampling negative examples from the entire set of items (i.e. easy negatives), choosing those that are closer to the positive samples as negatives (i.e. hard negatives) can help the model better distinguish between positive and negative samples \cite{ying2018graph}.
In graph structure, the distance of two nodes can reflect the similarity between them to a certain extent. With this in mind, for mining hard negative samples,
we propose a random-walk-based method. Specifically, for target node $n_{target}$,
% and his N-hop subgraph $(\mathcal{G},n_1,\cdots,n_k)$,
we perform $2^{nd}$ order random walk \cite{grover2016node2vec} starting from $n_{target}$. The unormalized transition probability $\alpha_{pq}(t,x)$ is defined as:
\begin{equation}
\alpha_{pq}(t,x)=
    \left\{
             \begin{array}{lr}
             \frac{1}{p}, & d_{tx}=0   \\
             1, & d_{tx}=1 \\
             \frac{1}{q}, & d_{tx}=2 
             \end{array}
\right.
\end{equation}
where the random walk just traversed edge $(t, v)$ and now at node $v$, $x$ denotes the nodes connected to $v$. Let's roughly estimate the depth (i.e. the longest distance to the target node) of random walk. If choose $x$ with $d_{tx}<=1$, such random walk approximate BFS behavior, only when $d_{tx}=2$ achieve DFS-like behavior. If whenever choosing $d_{tx}=2$, the shortest distance between the target node and $x$ strictly increases, we can approximate that this process follows a binomial distribution with $\hat{\emph{p}}=\frac{p}{p+pq+q}$. Thus, the upper bound on the expectation of walk depth $D$ from target node is:
\begin{equation}
    D \approx L\hat{\emph{p}}=L\cdot\frac{p}{p+pq+q},
\end{equation}
$L$ denotes the length of random walk. Intuitively, nodes farther from the target have less similarities with it, thus we sample N-hop neighbors and choose an
appropriate value $L$, such that $D$ is slighly larger than $N$, and take the last element from random walk sequence as hard negative, as described in Figure \ref{fig:hn_sampling}. Finally, we combine hard negatives with easy ones.

\subparagraph{\textbf{Total Objective Function.}} Objective (\ref{eq-multihop}) discriminates the target embedding in global latent space and objective (\ref{NIB-loss}) removes the information redundancy in local feature aggregation procedure. They are both critical and thus we add them to derive the final objective function of our SAC:
\begin{equation}
\label{final-loss-SAC}
\begin{split}
    \mathcal{L}_{SAC}=
    &\mathcal{L}_{Vanilla-SAC}+\eta\mathcal{L}_{NIB},
\end{split}
\end{equation}
where $\eta$ is the weight factor of NIB loss. Equipped with NIB, our SAC is able to learn more useful representations that optimally balance expressiveness and robustness of nodes in graphs, which is proved by extensive experimental results in next section.

\subsection{\textbf{Relation with Existing Works}} 
\label{relation_with_MLM}
With the success of Masked Language Modeling (MLM) \cite{devlin2018bert} in natural language processing area, many related efforts 
% related with MLM 
have been made in other areas including recommender systems.
In a recent work \cite{xiao2021uprec}, a BERT-like framework, UPRec, is proposed. Similar to MLM, it devises a Mask Item Prediction (MIP) task to embed the user by predicting the masked item from the historical user-item interactions sequence. First of all, MIP focuses on temporal domain modeling by only using first-order interactions, while SAC on spatial. 
In addition, although SAC and MLM
% -based method
 both build a prediction task for target embedding, the differences between them are still significant (also depicted in Fig.\ref{fig:f3}):
\begin{itemize}
    \item MLM-based methods mainly focus on optimizing the encoder network by masking and reconstructing the masked neighbor, which is an implicit and indirect way to optimize the embedding for target node. On the contrary, our SAC explicitly optimizes the target node embedding by 
    % masking and contrasting the masked neighbor, 
    minimizing the contrastive loss between the masked neighbors and the target,
    which simultaneously enables the encoder network to produce the neighbor-predictive coding conditioned on the surrounding neighbor context.      
    \item MLM-based methods perform feature aggregation with the masked neighbor, which is usually filled with 0 value or random value. We argue that it would additionally bring noises into the other embeddings as illustrated in \citet{he2021masked}. Conversely, our SAC does not take the masked neighbor as input and use its original vector as the positive sample for target node. Such paradigm could also explicitly optimize the embedding of the neighbour in the masked position, which benefits from the direct gradient flow from contrastive loss. 
    % \item UPRec collects the historical interacted items of user to generate the sequence while ignoring the high order proximity, focuses on modeling the temporal interactions. In SAC, we pay more attention to the spacial proximity and reconstructing the high-order connectivity.
\end{itemize}
To summarize, the target-to-neighbor contrastive paradigm is more effective than MLM-based ones, and SAC could explicitly optimize the embeddings of both the target node and the masked neighbors. Besides, our Multi-Hop Neighbor Modeling strategy further improves the optimization, also never been explored previously.

\section{Experiments}
We verify SAC on three public benchmark datasets as well as a web-scale dataset Douyin-Friend-Recommendation.
% which is for user recommendation. 
We seek to answer the following research questions:
\begin{itemize}
    \item \textbf{RQ1}: How does SAC perform compared with state-of-the-art methods especially GNN-based methods on public datasets?
    \item  
    \textbf{RQ2}: 
    % What's the influence of different hyper-parameter settings and different parts of SAC, e.g. the number of neighbor hops for sampling the target node's context, the number of nodes for each hop, the information bottleneck architecture.
    How different hyper-parameter settings and different parts of SAC influence the performance?
    \item \textbf{RQ3}: Is there any improvement in practical application scenarios, especially for long-tail user?
\end{itemize}

\begin{center}
\setlength{\abovecaptionskip}{0.1cm}
\begin{table}
\caption{Statistics of datasets.}
\label{table_data_statistics_benchmark}
    \begin{threeparttable}
  \begin{tabular}{ccccc}
  \toprule
                        & \#Users & \#Items & \#Interactions & Sparsity\tnote{1}  \\
    \midrule
  Book-Crossing         & 105283 & 340553 & 1149780 & $3.21*10^{-5} $ \\
  % 35854125650
   User-Behavior        & 966257 & 945443 & 19354068 & $2.12*10^{-5} $\\
   % 913540916851
   Yelp                 & 2189457& 160585 & 8635403 & $ 2.46*10^{-5}$ \\
   % 351593952345
     \bottomrule
  \end{tabular}  
  \begin{tablenotes}   
        \footnotesize               
        \item[1] Computed by $\frac{\#Interactions}{\#Users\times\#Items}$
      \end{tablenotes}  
  \end{threeparttable}
  \end{table}

\end{center}

\subsection{Experimental Setup}

\subsubsection{\textbf{Datasets}}

\subparagraph{\textbf{Benchmaerk Datasets.}} We use three public benchmark datasets Yelp\footnote{https://www.yelp.com/dataset}, User-Behavior\footnote{https://tianchi.aliyun.com/dataset/dataDetail?dataId=649} and Book-Crossing\footnote{http://www2.informatik.uni-freiburg.de/~cziegler/BX/} \cite{ziegler2005improving} to answer \textbf{RQ1} and \textbf{RQ2}. 
For Yelp and Book-Crossing, we directly generate the user-item bipartite graphs from the raw data\footnote{The number of items of Book-Crossing is 340553, which is different from the description on the website.}. For User-Behavior, 
% to evaluate the performance of baseline methods with public code 
in order to reproduce the baselines using public code as much as possible while avoiding memory overflow, we randomly select about 22\% items from the whole 4162024 items, and delete the users that have no interactions with any of these selected items.
For Yelp, we select the latest 20\% reviews for test, which are timestamped later than September 1, 2018, there are a total of 6620865 training samples and 2014538 testing samples. For Book-Crossing, we select 20\% interactions for each user to build test set, there are a total of 928389 training samples and 221391 testing samples. For User-Behavior, we leave the interactions later than December 2, 2017,12:00:00 for testing, there are a total of 14682436 training samples and 4671632 testing samples. We hold out 10\% interactions from the training set as validation set for all of these three datasets.
See Table \ref{table_data_statistics_benchmark} for statistics of these three datasets.

\subparagraph{\textbf{Douyin Dataset.}}We conduct extra experiments on Douyin-Friend-Recommendation to answer \textbf{RQ3}.
Douyin is a short-form-video-focused social networking service owned by Chinese company ByteDance Ltd.
We extract users' social relationships and the historical friend recommendation data to build a social graph, and produce representations for users with SAC then add to downstream friend recommendation models.
We build the graph by collecting users' social relationships and historical behaviors earlier than July 31st, 2021 from the database, including users' historical follow records, social relations. The graph is composed of 1.5 billion users and 250 billion edges. 
% Different from the first three public datasets, each user has thirty-three discrete features including gender, age range, location information, job information, etc. We turn these discrete features into learnable embeddings and add them to the user's ID embedding. 
We use historical recommendation data between 
August 14, 2021 and August 24, 2021 for evaluation. On average, there are 0.65 billion samples per day. 
\textbf{All of the used information including the user's ID and basic attributes, as well as the user's historical behavior records, have been strictly desensitized: all user IDs are hashed, and all attribute values are also anonymized, the user's relationship information does not contain any sensitive information}.
% See Appendix \ref{appendix_dataset} and \ref{appendix_douyin_datasets} for details of these datasets.

\subsubsection{\textbf{Implementation Details.}}
To make our model scalable to any scale, we implement SAC by Tensorflow and deploy it on the distributed architecture PS-Worker\footnote{PS: Parameter Server.}. We also use the graph engine euler\footnote{https://github.com/alibaba/euler} to build a real-time neighbor sampling service for performing neighbor query operation high efficiently. 
We implement all baseline models based on publicly available code. For LINE, HOP-Rec and PinSage, we directly implement the code in Tensorflow and deploy it on PS-Worker Architecture. For NGCF\footnote{https://github.com/huangtinglin/NGCF-PyTorch}, LightGCN\footnote{ https://github.com/gusye1234/LightGCN-PyTorch}, GCC\footnote{ https://github.com/THUDM/GCC}, we reimplement the code (we modify the data processing code to avoid memory overflow) based on their public PyTorch version  and run experiments on Linux servers with 256 RAM and NIVIDA RTX 3090 GPU. For SGL, we reimplement it with their public Tensorflow-based code \footnote{https://github.com/wujcan/SGL}.
\\
\subsubsection{\textbf{Settings.}}
To optimize the objective (\ref{final-loss-SAC}), following most existing works, for each positive pair, we sample 4096 easy negative nodes from $\mathcal{B}^-$ (defined in \ref{NCE_describe}),
and choose 16 hard negatives by performing $2^{nd}$ random walk with $p=1, q=0.5$, and select the last elements in each walk. We tune the length of walk within \{8, 10, 12, 14\}.
We choose Adagrad \cite{duchi2011adaptive} as the optimizer with learning rate $lr=0.001$. All of the parameters are initialized with the Xavier \cite{glorot2010understanding}. 
The batch size is fixed to 1024, the embedding size(also the hidden size) is 128 for Book-Crossing, and 256 for others.
we tune the Transformer encoder layers within the range of \{1, 2, 3, 4, 5, 6\}, 
We tune the weight factor $\eta, \beta, \tau$ in loss (\ref{NIB-loss}) and (\ref{final-loss-SAC}) within \{0.01, 0.05, 0.1, 0.2, 0.5, 1.0, 2.0\},  \{0.005, 0.01, 0.02, 0.05, 0.1, 0.5, 1.0\},  \{0,1, 0.2, 0.5, 1.0\} respectively,
the hop of neighbors for generating the node's context within the range of \{1, 2, 3, 4, 5\}, the number of heads in Attention mechanism within \{1, 2, 4, 8\}, the number of sampled neighbors $S_i$ at each hop $i$ in \{4, 8, 16, 32, 64, 128\} and follow the rule $\prod{S_i}<=512$, since nodes farther from the target node are noisier, we sample more neighbors for closer hops and less for farther hops.

\subsubsection{\textbf{Baselines}.} 

We compare our proposed SAC, with the following methods:
\begin{itemize}
    \item \textbf{LINE} \cite{tang2015line}: 
    % A classic large-scale graph embedding algorithm which consider both local similarity(first order proximity) and global similarity(second order proximity) to generate the node's representation. It also proposes an edge sampling method. LINE is similar to matrix factorization in embedding based recommendation.
    A classic large-scale graph embedding algorithm which consider both local and global similarity. We directly use the generated embedding for evaluation. We directly train LINE on the user-item bipartite graph.
    % \item \textbf{GraphSAGE} \cite{hamilton2017inductive}: Different from traditional GCNs learn the representation on a fixed graph structure(i.e. transductive), GraphSAGE proposes an inductive way, which can represent the nodes that do not appear during the training process by learning the  function that aggregates neighbors' information.
    % \item \textbf{NeuMF} \cite{he2017neuralNCF}: 
    \item \textbf{HOP-Rec} \cite{yang2018hop}:It combines MF and Graph-based models. Based on the original MF method, different types of positive samples are sampled with a certain probability through random walk on graph.
    \item \textbf{MLM}: Following Mask Item Prediction (MIP) proposed in UPRec \cite{xiao2021uprec}, we devise an MLM-based model, which uses the BERT as encoder and multi-hop neighbors as context, and produces the users' embedding by employing max-pooling operation on final hidden representations.
    % We aim at comparing MLM and SAC paradigm.
    \item \textbf{PinSage} \cite{ying2018graph}: As an variant of GraphSage \cite{hamilton2017inductive}, it generates node's context by random walk and proposes a hard negative sampling methods. It is the first work that deployed GCN to web-scale recommender systems.
    \item \textbf{NGCF} \cite{wang2019neural}: This is a CF method which performs message-passing progress over user-item bipartite graph to explicitly model the high-order connectivity.
    % The process of embedding propagation is similar to GCNs that iteratively aggregate neighbors' information. 
    \item \textbf{LightGCN} \cite{he2020lightgcn}: 
    % Based on NGCF, 
    It proposed that the linear transformation and non-linear activation in GCN do not have much effect on collaborative filtering, and even have side effects on the performance of recommendation. So LightGCN is only composed of neighborhood aggregation.
    \item \textbf{GCC} \cite{qiu2020gcc}: A graph self-supervised pre-training framework. It performs random walk and node anonymization to generate subgraphs 
    % and treats two subgraphs induced from the same local structure as a positive pair. 
    and optimizes InfoNCE to learn transferable structural representations. 
    We get the similarity between user and item by computing the dot product of embedding.
    \item \textbf{SGL} \cite{wu2021self}: Similar to GCC, this work also introduces self-supervised learning on graph. It uses node dropout, edge dropout and random walk to generate different views for nodes and maximizes the agreement between different views of the same node compared to that of other nodes.
\end{itemize}
It is worth noting that different from most existing works that verified their models 
on small-sized datasets which is a small subset of the original ones,
% which are small parts of the corresponding original datasets
\textbf{we select larger and sparser ones}. For example, 
% the sparsity of datasets used in \cite{he2020lightgcn,he2017neuralNCF} is ten times or even hundred times larger than ours, while 
the scale of datasets used in \cite{he2020lightgcn,he2017neuralNCF} is ten or even hundred times smaller than ours. \textbf{In addition, we directly use the raw data of Yelp and Book-Crossing without pruning}. Datasets with larger scale and sparser connections are more in line with real scenario, which can better verify the scalability of SAC.
% To validate the effective and scalability of SAC, we compare our proposed model with the following baselines: (1) \emph{Traditional Graph Embedding Methods}: including \textbf{LINE} \cite{tang2015line}, \textbf{Hop-Rec} \cite{yang2018hop}, 
%  (3) \emph{GNNs}: including \textbf{PinSage} \cite{ying2018graph}, \textbf{NGCF} \cite{wang2019neural}, \textbf{LightGCN} \cite{he2020lightgcn}. All of these models are delicately-designed for recommendation and follow a common GNN-based paradigm as described in section \ref{section_preliminary}.
% (4) \emph{Self-supervised GNNs}: Many works have shown the effectiveness of 
% combining graph-learning with the 
% self-supervised paradigm, also in the recommendation area \cite{qiu2020gcc, wu2021self,liu2021graph,hassani2020contrastive}. 
% Thus we select two representative works \textbf{GCC} \cite{qiu2020gcc} and \textbf{SGL} \cite{wu2021self} for comparison.

\subsubsection{\textbf{Evaluation Metrics}}

For each user, we treated all items 
% that he has not interacted with 
have no interaction with him as negative samples. We get the user's preference scores on all items by computing the dot product of the embeddings between them.
For \textbf{RQ1} and \textbf{RQ2}, We adopt two commonly used metrics Recall@k and NDCG@k(k=20 by default).
% In addition, we rank all items for each user to generate the top-k list but not compute the sampled metrics.
In addition, instead of estimating metrics by sampling, we rank all items for each user to generate a top-k list.
For real scenario friend recommendation task, we use AUC (Area Under the receiver operating characteristic Curve) 
and UAUC (User grouped AUC) 
for evaluation, we will describe more details in section \ref{expr_rq3}.

\begin{center}
\setlength{\abovecaptionskip}{0.1cm}
\begin{table*}
\caption{Overall Performance Comparison. The percentage in brackets denote the relative performance improvement over SGL.}
\label{table_overall_performance_comparison}
  \setlength{\tabcolsep}{1mm}{
  \begin{tabular}{c|cc|cc|cc}
  \toprule
                        \multirow{2}{*}{\textbf{Methods}} & \multicolumn{2}{c|}{\textbf{Book-Crossing}} & \multicolumn{2}{c|}{\textbf{User-Behavior}} & \multicolumn{2}{c}{\textbf{Yelp}} \\
                        \cline{2-7}
                        & \textbf{Recall} & \textbf{NDCG}  & \textbf{Recall} & \textbf{NDCG}  & \textbf{Recall} & \textbf{NDCG} \\
                        
    \midrule
    % \midrule
                        \large LINE & \large 0.0232 & \large0.0104& \large0.0147& \large0.0075& \large0.0070& \large0.0036\\
                        
                       \large HOP-Rec & \large0.0287& \large0.0155& \large0.0152& \large0.0087& \large0.0076& \large0.0034\\
                       \large  MLM &\large0.0308 & \large0.0173 & \large0.0148& \large0.0089 &  \large0.0071& \large0.0041\\
                      \large  PinSage &\large0.0315 & \large0.0172 & \large0.0179& \large0.0118&  \large0.0086& \large0.0045\\
                      \large  NGCF & \large0.0323 & \large0.0174& \large0.0157& \large0.0094& \large0.0077& \large0.0039\\
                     \large   LightGCN &  \large0.0387& \large0.0209& \large0.0178& \large0.0114& \large0.0082& \large0.0043\\
                     \large   GCC & \large0.0384 & \large0.0181& \large0.0182& \large0.0126& \large0.0118& \large0.0041\\
                     \large   SGL & \large0.0413 & \large0.0235& \large0.0188& \large0.0141& \large0.0124& \large0.0049\\
    
    %   SAC-c1 & 0.0413 & 0.0235& 0.0188& 0.141& 0.0124& 0.0049\\
    %   SAC-c1 & 0.0413 & 0.0235& 0.0188& 0.141& 0.0124& 0.0049\\
    %     SAC-c1 & 0.0413 & 0.0235& 0.0188& 0.141& 0.0124& 0.0049\\
    %      SAC-c1 & 0.0413 & 0.0235& 0.0188& 0.141& 0.0124& 0.0049\\
    \midrule
                        % \textbf{SAC(-NIB)} & 1 & 1& 1& 1& 1& 1\\
                  SAC (base)  & 0.0425 (+2.91\%)& 0.0246 (+4.68\%)& 0.0193 (+2.66\%)& 0.0144 (+2.13\%)& 0.0131 (+5.65\%)& 0.0052 (+6.12\%)\\
                  SAC (base) + \emph{mhop} & 0.0433 (+4.84\%)& 0.0251 (+6.81\%)&
                  \textbf{0.0201 (+6.91\%)}& \textbf{0.0152 (+7.80\%)}&
                   0.0134 (+8.06\%)& 0.0054 (+10.20\%)\\
                  SAC (base) + \emph{NIB} & \textbf{0.0439 (+6.30\%)}& \textbf{0.0254 (+8.09\%)}& 0.0198 (+5.32\%)& 0.0149 (+5.67\%)& \textbf{0.0135 (+8.87\%)}& \textbf{0.0056 (+14.29\%)}\\
                  SAC (base) + \emph{hn} & 
                  0.0431 (+4.36\%)& 0.0249 (+5.96\%)& 0.0196 (+4.26\%)&
                  0.0146 (+3.55\%)& 0.0131 (+5.65\%)& 0.0053 (+8.16\%)\\
                  \midrule
                     \large   \textbf{SAC} & \textbf{0.0446 (+7.99\%)} & \textbf{0.0257 (+9.36\%)} & \textbf{0.0205 (+9.04\%)} & \textbf{0.0155 (+9.93\%)} & \textbf{0.0139 (+12.10\%)} & \textbf{0.0058(+18.37\%) }\\

     \bottomrule
  \end{tabular}  
  }
%   \vspace{-0.5cm}
  \end{table*}
\end{center}

% \begin{center}
% \setlength{\abovecaptionskip}{0cm}
% \setlength{\belowcaptionskip}{0cm}  
% \begin{table}
% \caption{Ablation study on SAC, evaluation result of Recall}
% \label{table_ablation_study}
%   \setlength{\tabcolsep}{1mm}{
%   \begin{tabular}{cccc}
%   \toprule
%     \textbf{Variants}& \textbf{Book-Crossing} & \textbf{User-Behavior} & \textbf{Yelp} \\
 
%     % \midrule
%     % \midrule
%     \midrule[0.7pt]
%     % MLM  & 0.0149 & 0.0053  & 0.0017 \\
%     SGL & 0.0413 & 0.0188 & 0.0124
%     \\
%     \midrule[0.4pt]
% % SAC-\emph{mhop} & 0.0378 & 0.0183& 0.0115\\
% %  SAC-\emph{hn} & 0.0423 & 0.0189& 0.0132\\
% %  SAC-\emph{NIB} & 0.0395 & 0.0192& 0.0124\\
% %  SAC-\emph{TF} & 0.0316 & 0.0163& 0.0081\\
% %     \midrule
% %  SAC & 0.0446 & 0.0205& 0.0139\\
% % SAC (MLP) & 0.0381 & 0.0171& 0.0094 \\
% SAC (base) & 0.0425 & 0.0193& 0.0131 \\
% SAC (base) + \emph{mhop} & 0.0433 & 0.0198 & 0.013 \\
% SAC (base) + \emph{NIB} & 0.0439 & 0.0201 &  0.0134\\
% SAC (base) +\emph{hn} & 0.0431 &  0.0196 &  0.0131\\

% % SAC (single-hop)\\
% % +mhop\\
% % +NIB\\
% % +hn\\

%      \bottomrule
%   \end{tabular}  
%   }
%   \vspace{-0.5cm}
%   \end{table}
% \end{center}

\subsection{Performance Comparison (RQ1)}
We compare the performance of SAC with all baseline methods. The evaluation results of different methods are shown in Table \ref{table_overall_performance_comparison}, from the top half and the last row of which we have several observations:
\begin{itemize}
    \item All of the neural-based methods consistently outperform LINE in three datasets,
    proving that 
    % the representations generated by traditional graph embedding methods in which 
    % only first-order information is directly used 
    merely using first-order information
    is not sufficient to capture the complex similarity between items and users.
    \item Hop-Rec and MLM performs better than LINE, but  there is still a significant gap compared with other GNNs, the performance of MLM implies that indirectly optimizing the target user's representation is not sufficient enough.
    % \item The improvements of Hop-Rec compared with LINE illustrate that introducing hight order connection to enrich the information of representation is effective while in LINE, we only use first order adjacency of user-item pair.
    \item 
    % PinSage performs basically the same with Hop-Rec in Book-Crossing, and yields improvements in User Behavior and Yelp. PinSage is an inductive GNN method which can also exploit high-order neighbors. It implies that in most cases, inductive GNNs have stronger capabilities than MF in modeling user-item interactions especially on larger and sparser graphs. 
    PinSage outperforms NGCF and LightGCN in User-Behavior and Yelp, while underperforms these two in Book-Crossing. Book-Crossing is much smaller that the other two.
    % PinSage is designed for Large sacle graph, it performs neighbor aggregating by sampling but not matrix multiplying.
    That implies transductive GNNs will be less effective on large-scale graphs compared with on smaller ones.
    % that implies not only transductive GNNs is costly on large graphs, but also it may also have negative impact on performance, that may because exploding number of neighbors brings a lot of noise.
    % \item 
    % Both NGCF and LightGCN achieves better performence than PinSage.  All of these three methods are GNN-based, this result shows that combining different layers' representation is better than only use the last layer's output. In addition, 
    LightGCN performs better than NGCF, verifying the non-linear transformation and weight matrix is useless in CF-based recommendation.
    \item GCC and SGL exploit self-supervising to enhance the representation learned by GNNs. 
    % We directly use the representations generated by GCC to evaluate the model's performance on recommendation.
    We find GCC performs closely to LightGCN in Book-Crossing and outperforms LightGCN in User-Behavior and Yelp. SGL consistently achieves the best result along all baselines on all datasets. The performance of both GCC and SGL proves that self-supervised paradigm for local structure reconstruction is a better choice for large-scale sparse graph embedding learning.
    \item SAC consistently yields the best performance over all of the three datasets.
    Specifically, for Recall@20, the improvement of SAC over SGL is 7.99\%, 9.04\%, 12.10\% in Book-Crossing, User-Behavior, Yelp respectively, for NDCG@20, the improvement is 9.36\%, 9.93\%, 18.37\%.
    There are similarities between GCC/SGL and SAC, both self-supervising and auto-regressive paradigm aims at learning 
    % and distinguishing 
    local topology by constructing subgraphs.
    The improvements of SAC over GCC/SGL shows the benefits of directly aggregating multi-hop neighbors.
    The improvements are more significant on User-Behavior and Yelp, which have sparser 
    connections between users and items, 
    implies
    multi-hop masking bring benefits for large-scale recommendation. 
    In addition, the improvements over traditional GNNs means SAC can efficiently select useful signals among a large number of neighbor information containing noise.
\end{itemize}

\begin{figure}
%  \vspace{1cm}  %调整图片与上文的垂直距离
 \setlength{\abovecaptionskip}{0.2cm}   %调整图片标题与图距离
  \setlength{\belowcaptionskip}{-0.4cm}   %调整图片标题与下文距离
% \subfigure[\textbf{Book-Crossing}]{
%   \begin{minipage}[t]{0.3\linewidth}
%     \centering
%     \includegraphics[scale=0.15]{fig_diff_num_hops_BookCrossing_0113.pdf}
%   \end{minipage}%
%   }
%   \subfigure[\textbf{User-Behavior}]{
%   \begin{minipage}[t]{0.3\linewidth}
%     \centering
%     \includegraphics[scale=0.15]{fig_diff_num_hops_UserBehavior_0113.pdf}
%   \end{minipage}
%   }
    {
  \begin{minipage}[t]{0.9\linewidth}
    \centering
    \includegraphics[scale=0.14]{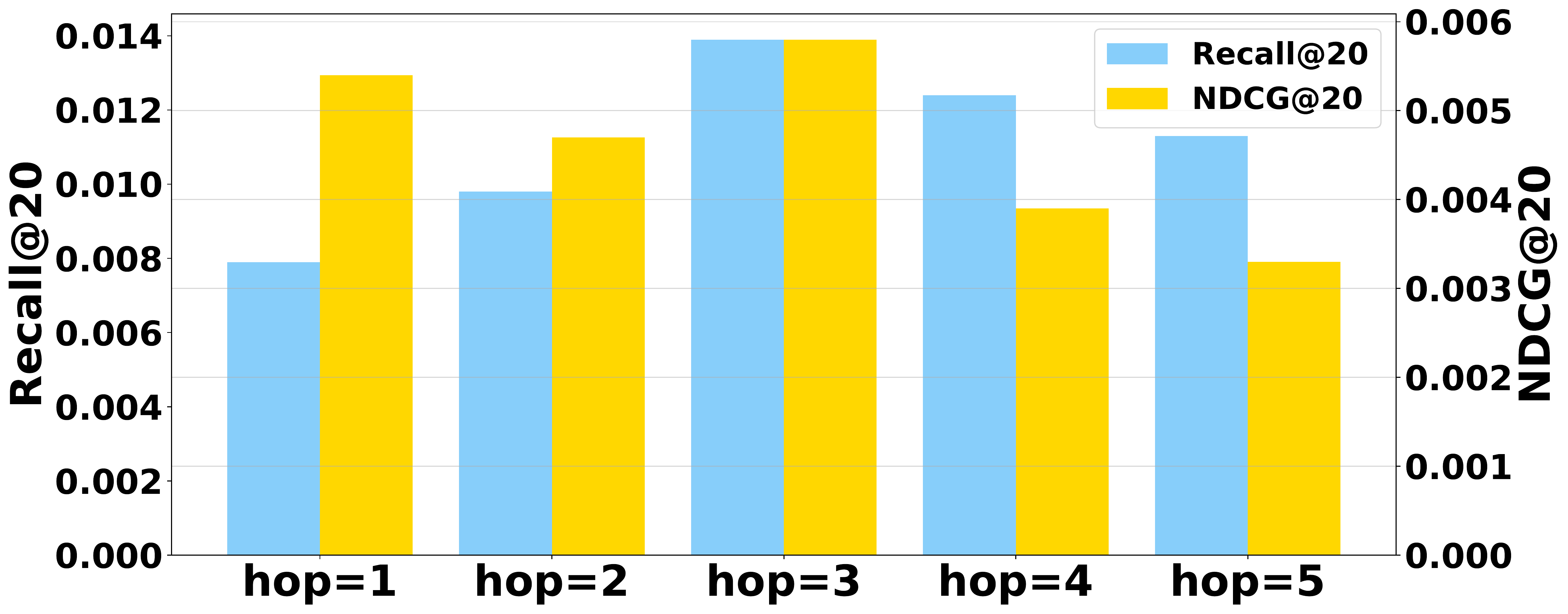}
  \end{minipage}%
  }
%   \subfigure[]{
%   % yelp
%   \begin{minipage}[t]{0.3\linewidth}
%     \centering
%     \includegraphics[scale=0.15]{fig_diff_num_hops_Yelp_0113.pdf}
%   \end{minipage}
%   }
   \caption{Performance of SAC w.r.t different number of hops for generating the context on Yelp.}
   \label{fig:dif_num_hops}
\end{figure}

\subsection{Model Analysis (RQ2)}
In this section, we study how different components and hyper-parameter settings affect the performance of SAC.
% \begin{figure}
%  \vspace{0cm}  %调整图片与上文的垂直距离
% \setlength{\abovecaptionskip}{0cm}   %调整图片标题与图距离
% \setlength{\belowcaptionskip}{-0.5cm}   %调整图片标题与下文距离
%   \subfigure{
%   \begin{minipage}[t]{0.9\linewidth}
%     \centering
%     \includegraphics[scale=0.20]{fig_ablation_study_0113.pdf}
%   \end{minipage}%
%   }
%     \caption{Ablation study: the contribution of different components:-c1: without multi hop masking; -c2: without NIB; -c3: without Transformer encoder.}
%     \label{fig:ablation_study}
% \end{figure}
\subsubsection{\textbf{Overall Ablation Study}}
There are three key components of SAC: (1) Multi-Hop Neighbor Modeling mechanism. (2) Neighbor Information Bottleneck (NIB).
% for model optimization.
(3) Hard negative sampling strategy.
To assess the impact of these modules,
% Besides, we use the Transformer as encoder.
we compare different ablation variants of SAC:
% by removing the above components respectively: In SAC-\emph{mhop}, we only mask one-hop neighbors. In SAC-\emph{hn}, we only sample easy negatives.
% In SAC-\emph{NIB}, we replace the loss function 
% (\ref{final-loss-SAC}) with the vanilla one (\ref{eq-multihop}). In SAC-\emph{TF}, we replace Transformer encoder with linear transformation and mean pooling. In addition, we have summarized the differences between SAC and MLM paradigm in section \ref{relation_with_MLM}, to verify the superior of SAC, we also conduct experiment on a variant with MLM paradigm, equipped with all four key components of SAC. 
In SAC (base), we remove all of the three key components (by removing (1), we only mask one-hop neighbors; by removing (2), we replace the loss function with the vanilla one; by removing (3), we only sample easy negatives). SAC (base)+\emph{mhop},  SAC (base)+\emph{NIB},  SAC (base)+\emph{hn} represent equipping SAC (base) with Multi-Hop Neighbor Modeling, NIB, hard negative sampling strategy respectively. 
% To verify the superior of SAC, we also conduct experiment on a variant with MLM paradigm, equipped with all three key components and Transformer encoder. 
The lower part of Table \ref{table_overall_performance_comparison} shows the experimental results, from which we have the following observations:

\begin{itemize}
    \item SAC (base) outperforms SGL and all other GNN-based methods, also including the MLM model equipped with Transformer, \textbf{proving the efficiency of spatial-autoregressive paradigm over GNNs}. Combining spatial-autoregressive paradigm with the Transformer, we equip the model with powerful global interaction capabilities, helps better capture local topology and high order proximity. 
    
    \item All of the three components have a positive effect 
    % on the performance 
    and combining them reaches a better result.
    By introducing multi-hop masking, we force SAC to model the high-order proximity, which is similar to augmenting the interactions between users and items, mitigates the negative effects of sparse interactions. We also find it has the greatest influence on User-Behavior, which is the sparest one, meaning the importance of modeling high-order proximity delicately.
    NIB having the greatest on both Book-Crossing and Yelp,  and it is also very significant on User-Behavior. The second part of loss (\ref{NIB-loss}) strictly limits the mutual information between predictive coding and neighbors, forcing the model to extract useful information. 
    The improvement of equipping hard negative sampling demonstrates the model can better  distinguish between positive and negative samples. 
    % \item Combining all components can reach a better result.
    
\end{itemize}

\subsubsection{\textbf{Impact of Neighbor Sampling Parameters}}

\begin{figure}[ht]
%  \vspace{-0.2cm}  %调整图片与上文的垂直距离
 \setlength{\abovecaptionskip}{0cm}   %调整图片标题与图距离
\setlength{\belowcaptionskip}{-0.1cm}   %调整图片标题与下文距离
  \subfigure[]{
  \begin{minipage}[t]{0.48\linewidth}
    \centering
    \label{fig:numbeer_sampled_nmodes_per_layer}
    \includegraphics[scale=0.137]{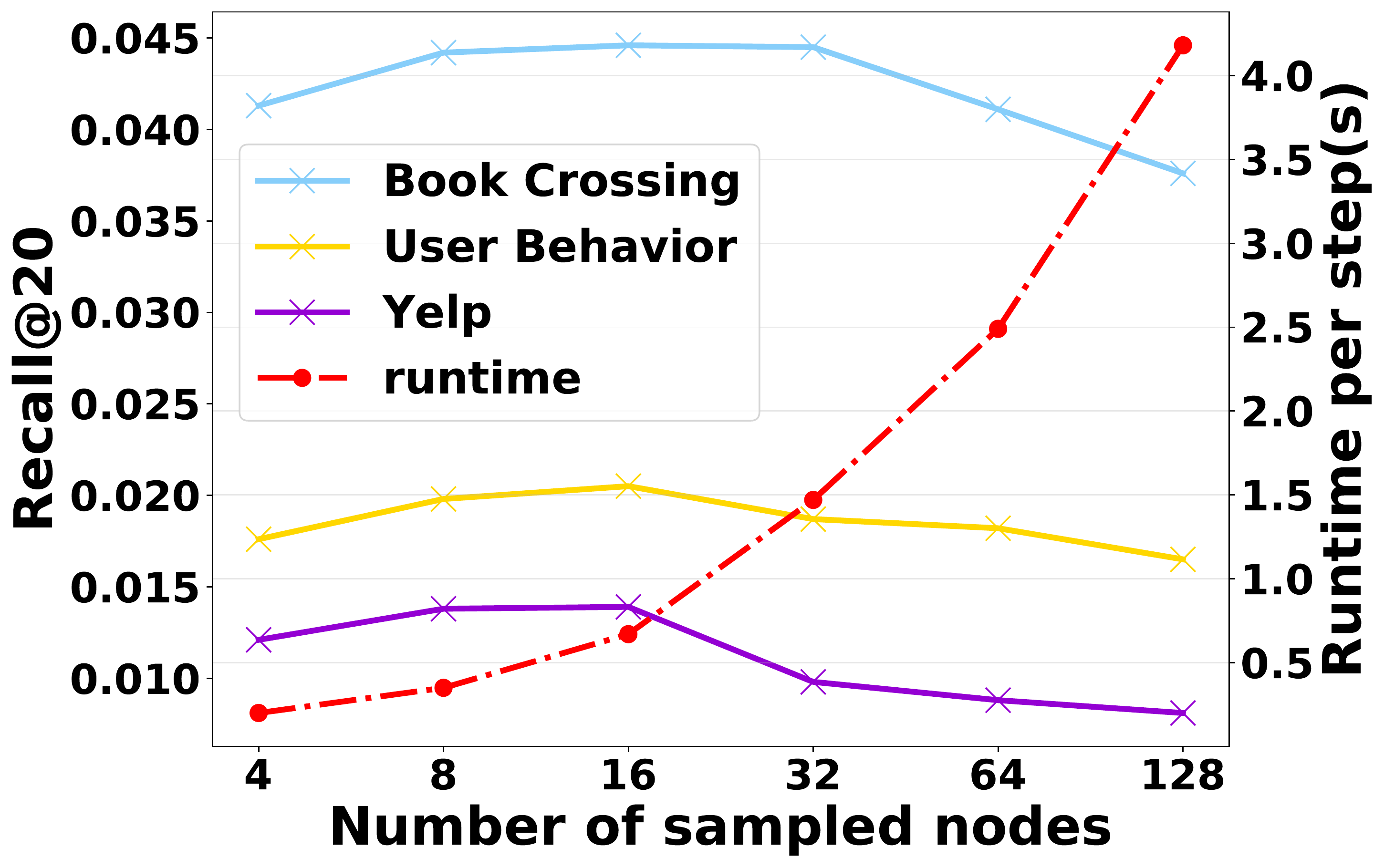}
  \end{minipage}%
  }
   \subfigure[]{
  \begin{minipage}[t]{0.48\linewidth}
    \centering
    \label{fig:numbeer_stacked_tf_layers}
    \includegraphics[scale=0.137]{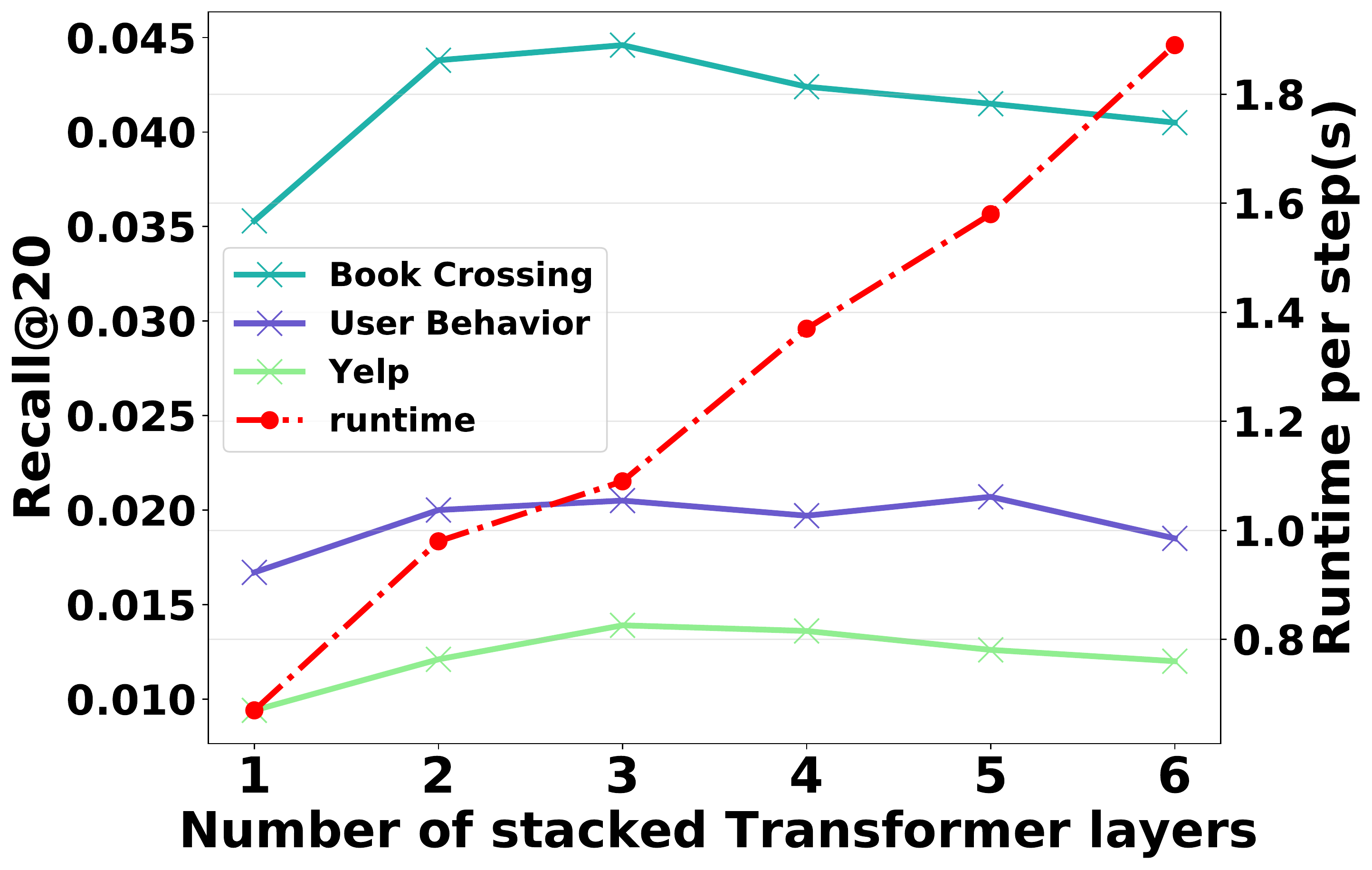}
  \end{minipage}%
  }
    \caption{Performance of SAC w.r.t different number of sampled neighbors and different number of Transformer layers.}
    % \label{fig:dif_num_neighbors_per_hop}
\end{figure}

% One of the most important operation in our spatial autoregressive paradigm is building the subgraph of the target node $n_{target}$. The spatial autoregressive coding $\mathbf{c}_p$ encode the neighbor proximity of subgraph into target node, the subgraph decides the scale of target node's context, in other words decides how many information the target node will aggreagte. 
We study how do neighbor aggregating parameters influence the model's performance.
There are two key parameters for generating the subgraph: (1). 
$hop$: How many hops will be considered to generate the subgraph. (2). $S_i$: How many number of nodes will be sampled at hop $i$. we conduct ablation study by setting $hop$ and $S_i$ to different values respectively. We search $hop$ in the range of \{1, 2, 3, 4, 5\}, $S_i$ in \{4, 8, 16, 32, 64, 128\}. For $S_i$, we set $i=1$. The experimental results are shown in Figure \ref{fig:dif_num_hops} and \ref{fig:numbeer_sampled_nmodes_per_layer}, we observe that: (1) Only using one hop neighbors 
    % to build the subgraph 
    degrades the performance compared with $hop=2$ and $3$. Intuitively, it is equal to merely use the observed interactions, totally ignoring useful high-order information. 
    % especially when the interactions is extremely sparse.
    When $hop \geq 3$, the performance begins to decline. Recap that we sample a fixed number of nodes at each hop, the number of noises grows fast with the size of the subgraph, bringing more biases. (2) Figure \ref{fig:numbeer_sampled_nmodes_per_layer} shows a similar result. The running time of each step grows rapidly with the number of sampled nodes at each hop, 
    which do not always achieve improvement of performance. An appropriate neighbor size (i.e. 16 to 32 in our experimental results) is important to achieve better generalization performance.
% \begin{itemize}
%     \item Only using one hop neighbors 
%     % to build the subgraph 
%     degrades the performance compared with $hop=2$ and $3$. Intuitively, it is equal to merely use the observed interactions, totally ignoring useful high-order information. 
%     % especially when the interactions is extremely sparse.
%     When $hop \geq 3$, the performance begins to decline. Recap that we sample a fixed number of nodes at each hop, the number of noises grows fast with the size of the subgraph, bringing more biases.
%     \item Figure \ref{fig:numbeer_sampled_nmodes_per_layer} shows a similar result. The running time of each step grows rapidly with the number of sampled nodes,
%     % at each hop, 
%     which do not always achieve improvement of performance. An appropriate neighbor size (i.e. 16 to 32 in our experimental results) is important for a better generalization.
%     % since too large number of nodes at one hop will make the number of nodes outside this hop explode, bring a lot of noise. 
% \end{itemize}
% \begin{figure}
%  \vspace{0.2cm}  %调整图片与上文的垂直距离
% \setlength{\abovecaptionskip}{0cm}   %调整图片标题与图距离
% \setlength{\belowcaptionskip}{-0.5cm}   %调整图片标题与下文距离
%   \subfigure[\textbf{Recall@20}]{
%   \begin{minipage}[t]{0.48\linewidth}
%     \centering
%     \includegraphics[scale=0.145]{fig_diff_TF_layers_recall_0113.pdf}
%   \end{minipage}%
%   }
%   \subfigure[\textbf{NDCG@20}]{
%   \begin{minipage}[t]{0.48\linewidth}
%     \centering
%     \includegraphics[scale=0.145]{fig_diff_TF_layers_ndcg_0113.pdf}
%   \end{minipage}%
%   }
%     \caption{Performance of SAC w.r.t different number of Transformer layers}
%     \label{fig:transformer_layer_nums}
% \end{figure}

\subsubsection{\textbf{Impact of the number of Transformer layers}}

We use Transformer to
% encode multi-hop information for the target node and 
produce neighbor predictive coding $\mathbf{c}_p$, We study the influence of the Transformer layers on the performance. From Figure \ref{fig:numbeer_stacked_tf_layers}, 
% we have the following observations:
% \begin{itemize}
%     \item Only a single encoder layer is not sufficient to modeling the high-order proximity of the subgraph. 
%     \item Stacking too much layers(greater than 4) will not significantly improve the model's performance, even have negative effect.
%     \item The forward propagation time of each step increases roughly linearly with the number of layers. In natural language, each sentence has a complex syntactic structure, requires more layers of transformers to extract features of different granularities. But for graph structure, the relative position to the target node can describe most of the information, which implies stacking 2 layers maybe enough. Thus too much layers are redundant.
% \end{itemize}
we find that only a single encoder layer is not sufficient to model high-order proximity. However, stacking too much layers (greater than 2) will not significantly improve the model's performance, even have negative effect, while forward propagation time of each step increases roughly linearly with the number of layers. For graph structure, the relative position to the target node can describe most of the structural information, which is different from natural language. Thus stacking too much layers is useless.

\subsubsection{\textbf{Training Curves Analysis}}
\begin{figure}[t]
    % \vspace{0.3cm}  %调整图片与上文的垂直距离
 \setlength{\abovecaptionskip}{0cm}   %调整图片标题与图距离
  \subfigure
%   [\textbf{Recall, Book-Crossing}]
  {
  \begin{minipage}[t]{0.9\linewidth}
    \centering
    \includegraphics[scale=0.135]{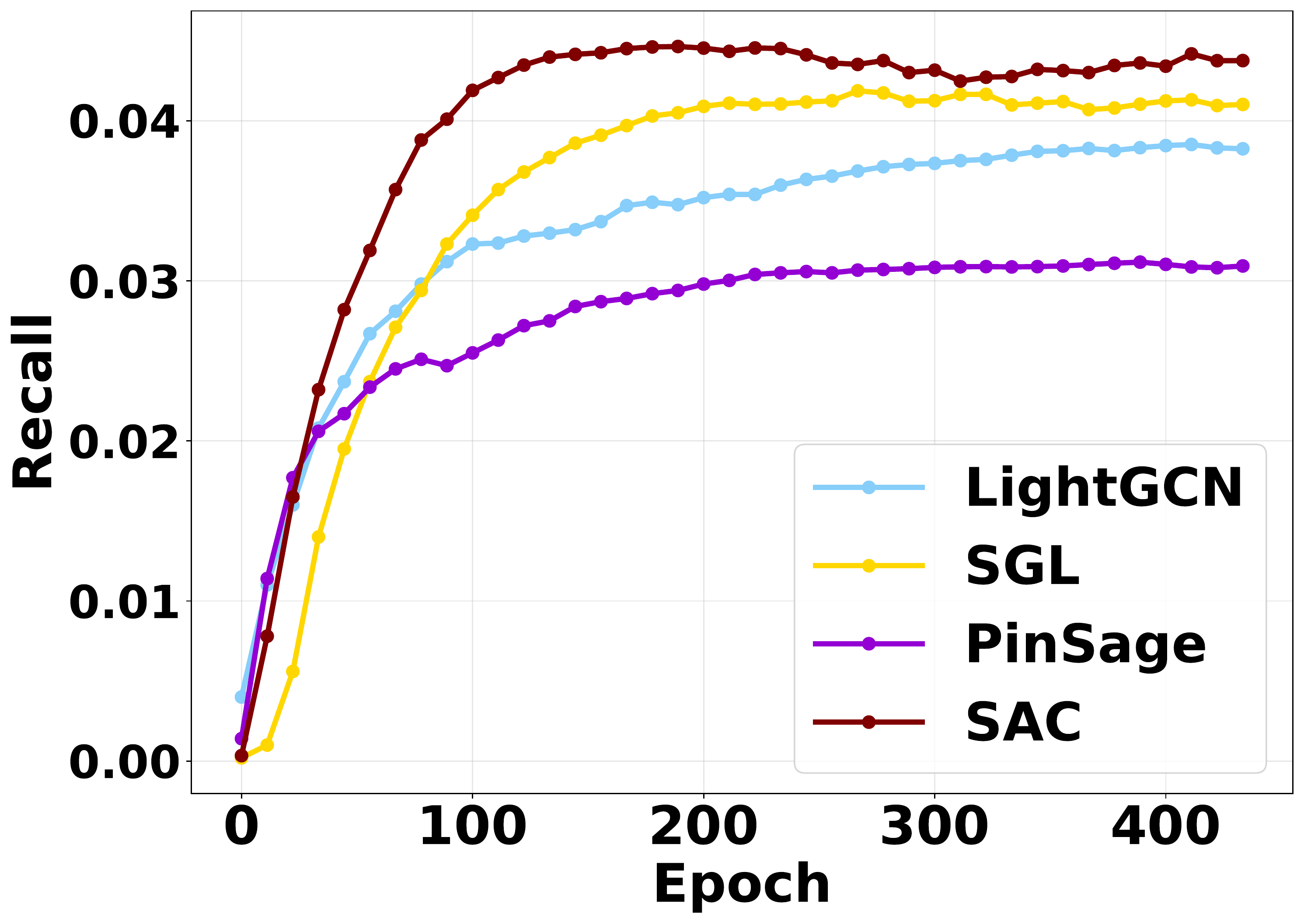}
  \end{minipage}%
  }
%   \subfigure[\textbf{MRR, Book-Crossing}]{
%   \begin{minipage}[t]{0.48\linewidth}
%     \centering
%     \includegraphics[scale=0.16]{fig_MRR_epoch.pdf}
%   \end{minipage}%
%   }

    \caption{Training curves of SAC and baselines on Book-Crossing: Changes in Recall with the number of epochs. }
    \label{fig3:training_curves}
\end{figure}

We study the trainning process of SAC compared with some baselines, Figure \ref{fig3:training_curves} shows the training curves of these models on Book-Crossing.
the performance on the other two datasets are similar. 
We find self-supervising and autoregressive paradigms (i.e. SGL and SAC) achieve faster converge speed compared with the other two models, and SAC converges faster than SGL. 
% \begin{itemize}
%     \item Self-supervised paradigm achieve faster converge speed compared with the other two models.
%     \item As the number of epochs grows, the other three baselines fell into overfitting earlier, while SAC can reach a better result.
% \end{itemize}
We attribute the speedups to the following highlights:
(1) Our spatial autoregressive paradigm is similar to masked language model, which has proved its superiority in natural language processing. But different from simply masking a node and predicting it, we exploit spatial autogregressive to generate the predictive embedding from target node and predict multi-hop neighbors at once, which enables the model to learn more information once. (2) Multi-hop masking make the target node \textbf{interact} with more nodes (e.g. items, users), which alleviates connection sparsity, thus implicitly alleviate overfitting caused by sparse interaction.

\begin{figure}
 \vspace{-1em}  %调整图片与上文的垂直距离
\setlength{\abovecaptionskip}{0cm}   %调整图片标题与图距离

   \subfigure[\textbf{daily AUC improvement}]{
  \begin{minipage}[t]{0.48\linewidth}
    \centering
    \includegraphics[scale=0.13]{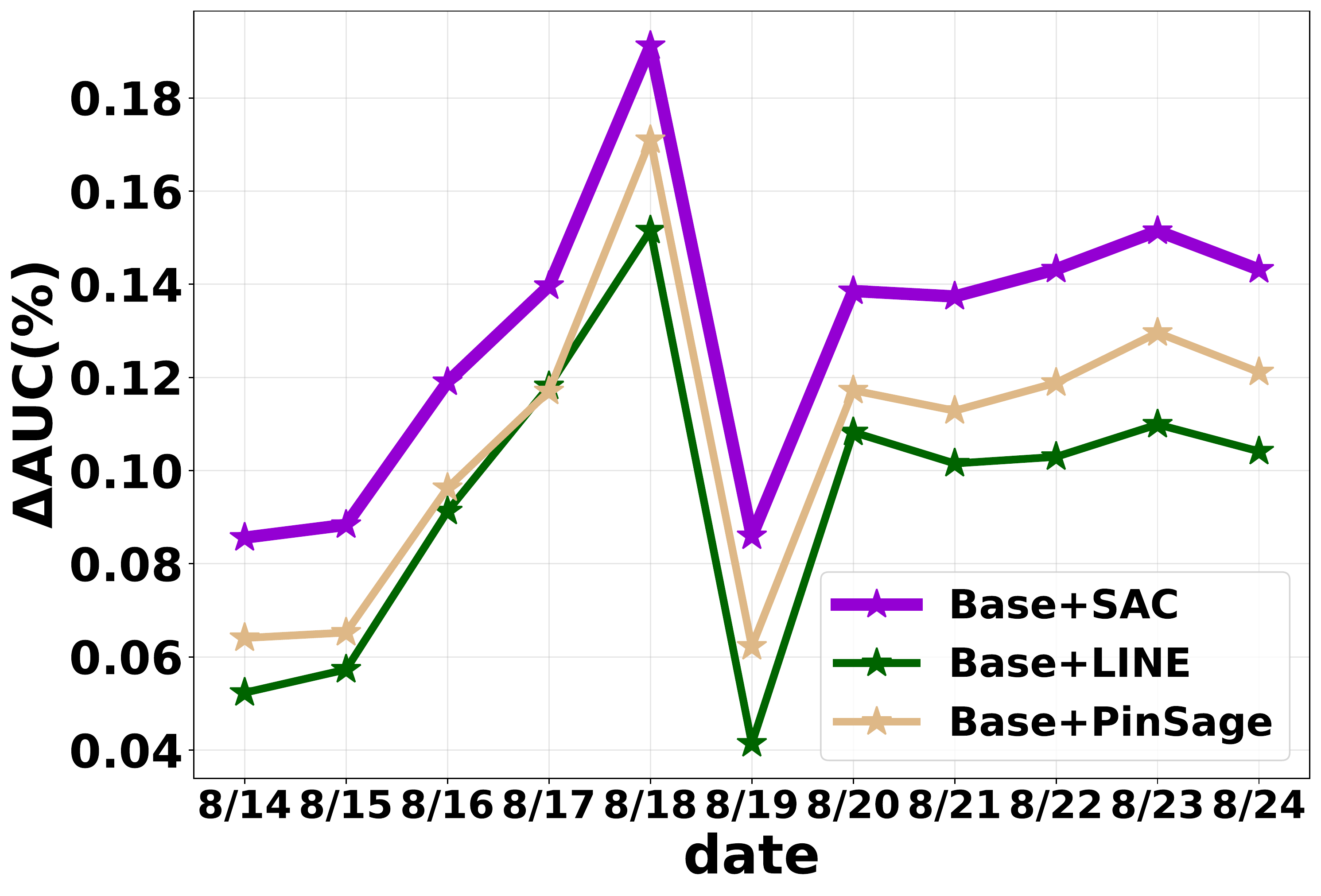}
    \label{fig:douyin_expr_offline_a}
  \end{minipage}%
  }
  \subfigure[\textbf{daily UAUC improvement}]{
  \begin{minipage}[t]{0.48\linewidth}
    \centering
    \includegraphics[scale=0.13]{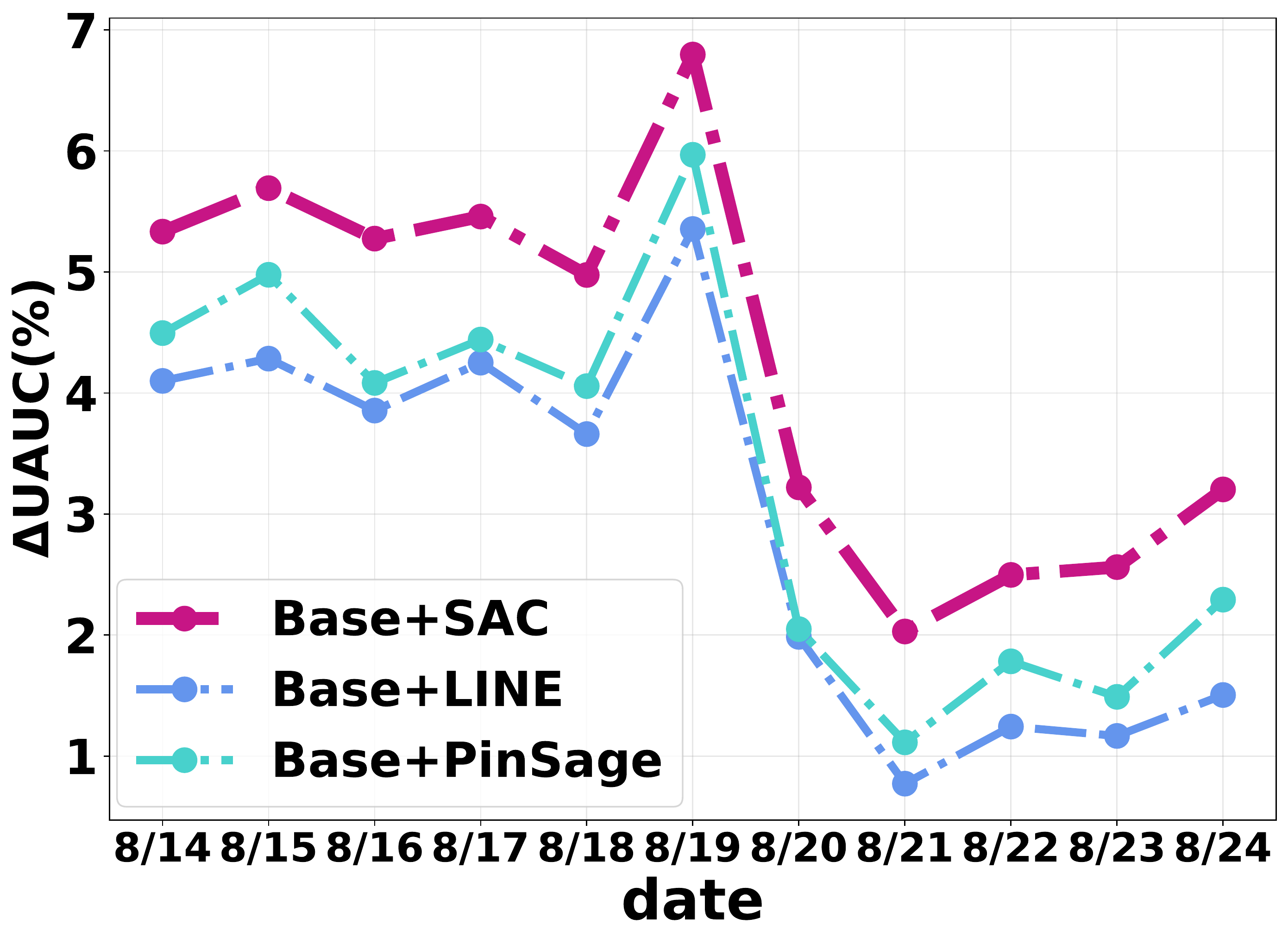}
    \label{fig:douyin_expr_offline_b}
  \end{minipage}%
  }
    \caption{The improvement of AUC (left) and UAUC (right) of different competitors compared with the Base model on Douyin-Friend-Recommendation.}
    \label{fig:douyin_expr_offline}
\end{figure}

\subsection{Evaluation on Online Recommendation Task (RQ3)}
\label{expr_rq3}

% to verify the quality of embedding generated by SAC, we train SAC on a web-scale graph of Douyin with 1.5 billion nodes, and evaluate performance on user recommendation task, 
To further verify the effectiveness of SAC in practical, we train it on a large-scale dataset Douyin-Friend-Recomme-ndation, in which the social graph is composed of  1.5 billion nodes, and evaluate the quality of representation by adding it to the downstream recommendation model,
which recommend other users for target users in their feed, that is similar to product recommendation. 
\\
\subparagraph{\textbf{Experiment Setup.}} 
% We build the graph by collecting users' social relationships and historical behavior from database. It contains 1.5 billion users and 250 billion edges. Different from the first three public datasets, each user has thirty three discrete features including gender, age range, location information, job information, etc. We turn these discrete features into learnable embeddings and add them to the user's ID embedding. We use historical recommendation data between 2021.8.14 and 2021.8.24 for evaluation. On average, there are 0.65 billion samples per day. 
% During evaluation, we select top-K nearest neighbors of the target user in embedding space for recommendation.
We conduct experiments with several competitors: (i) \emph{Base}: Two-tower-based model. (ii) \emph{Base+LINE}: Base model with embedding features generated by LINE. (iii) \emph{Base+
PinSage}: Base model with embedding features generated by PinSage.
(iv) \emph{Base+SAC}: Base model with embedding features generated by SAC.
% \\
% \subparagraph{\textbf{Metrics.}} 
% We focus on the following two metrics that are more practical in real scenario application: 
We focus on the following two practical metrics:
(i) AUC: we treat (user, recommended user) pairs which have real interactions as positive samples.
% , in which the user followed the recommended user.
(ii) UAUC: Different from AUC which sorts all samples according to the predicted value to calculate the AUC, UAUC is 
obtained by calculating the AUC of each user respectively, and takes the average of them. 
The UAUC can better reflect the quality of recommendation from user's perspective.
\\

\subparagraph{\textbf{Improvement for Friend Recommendation.}} Figure \ref{fig:douyin_expr_offline} shows the evaluation result of 
daily recommendation 
between  August 14, 2021 and August 24, 2021. 
We find 
% that it is useful to generating representations for users on large-scale social graphs and adding them to downstream recommendation models can improve recommendation metrics,
adding the embedding to downstream recommendation model 
% is useful, 
% both LINE and SAC 
can gain a positive growth of AUC and UAUC, SAC achieves more impressive results than LINE and PinSage. SAC increase the 
AUC by more than 1 thousandths on average. In real scenarios, 1 thousandths increment is already a significant improvement. For UAUC, we can see 
% that introducing pre-trained embedding to the recommendation model can reach 
a more significant improvement. Replacing LINE and PinSage with SAC can amplify the improvement, up to about 7\%. On average, the improvement of SAC compared with LINE is more than 1\%, which means the quality of recommendation in user's feed has improved a lot.
\\

\subparagraph{\textbf{Long-tail User Recommendation}}. At last, we study the effectiveness of SAC for long-tail users. We split all users into three groups based on the number of friends, fewer friends means the user has sparser social interactions, recommendation quality for this group of users reflects the capability for modeling long-tail users. From Table \ref{table_group_user_auc}, we find about 40\% of users have less than 60 friends, which once again confirms the long tail phenomenon 
% in large-scale recommendations.
The evaluation results 
demonstrate that
% those with low number of friends occupy a considerable part of all users. 
compared with LINE and PinSage, SAC significantly improves the recommendation performance on long-tail user. From both Figure \ref{fig:douyin_expr_offline} and Table \ref{table_group_user_auc}, we can conclude that 
% in this user recommendation task,
the improvement using SAC is mostly attributed to the improvement of recommendation performance on long-tail user.
{
\begin{table}[]
\caption{The Statistics of different user groups and the improvement comparision over different user groups.}
\label{table_group_user_auc}
% \large
% \begin{tabular}{c|c|c|c}

% \midrule
% Methods         & \multicolumn{3}{c}{delta AUC (\%)}                                \\ \hline
% Base+LINE            & 0.16                     & 0.34                      & 0.28  \\ \hline
% Base+PinSage         & 0.22                     & 0.47                      & 0.42  \\ \hline
% \textbf{Base+SAC}   & \textbf{0.27}                     & \textbf{0.70}                      & \textbf{0.69}  \\ \hline
% \hline
% User Group      & \multicolumn{1}{c|}{all} & \multicolumn{1}{c|}{0-5}  & 0-60  \\ \hline
% \% of all users & \multicolumn{1}{c|}{100} & \multicolumn{1}{c|}{9.54} & 37.36 \\ 
% \midrule
% \end{tabular}
\small
\begin{tabular}{l|l|l|l||l|l}
\hline
Methods                    & {\begin{tabular}[c]{@{}c@{}}LINE\end{tabular}} & {\begin{tabular}[c]{@{}c@{}}PinSage\end{tabular}}  & {\begin{tabular}[c]{@{}c@{}}\textbf{SAC}\end{tabular}}  & User Group & \% users \\ \hline
\multirow{3}{*}{\begin{tabular}[c]{@{}c@{}}$\Delta$AUC \\ (\%)\end{tabular}} & 0.16      & 0.22         & \textbf{0.27}     & all        & 100             \\ \cline{2-6} 
                           & 0.34      & 0.47         & \textbf{0.70}     & $\leq$5 friends       & 9.54            \\ \cline{2-6} 
                           & 0.28      & 0.42         & \textbf{0.69}     & $\leq60$ friends       & 37.36           \\ \hline
\end{tabular}
\end{table}
% \vspace{-1em}
}

\section{Conclusion}
In this work, we propose Spatial Autoregressive Coding (SAC), a unified novel framework for graph neural recommendation. Different from conventional multi-pass graph signal propagation which is sensitive to over-smoothing problems,
% to aggreagte multi-hop neighbors,
SAC straightly aggregates the high-order node context via spatial autoregressive paradigm, and then approximates diverse masked neighbors in a contrastive fashion to make full use of the connectivity information. 
% To directly produce embedding for recommendation task, 
In addition, a negative sampling strategy is specially designed to boost the model's capability of distinguishing between positive and negative samples.
We also introduce a Neighbor Information Bottleneck which suppresses redundancy and potentially adverse noise to optimize the reliability of compact graph substructure representations.
% which reconstructs the connectivity of user and items by absorbing high order proximity and predicting multi-hop links, while avoids excessive noise. 

Through performing quantitative and qualitative analyses on both medium- and large-scale benchmarks as well as Douyin online dataset, we probe the effectiveness of our proposed SAC, and set forth the applicability of autoregressive paradigm in practical graph-based recommendation systems. 

%We initially attempt to exploit autoregressive into this field. We would like to go deeper into the application of the autoregressive paradigm in recommender systems and explore more frameworks for recommendation and other tasks.

% \section{Appendices}

% If your work needs an appendix, add it before the
% ``\verb|\end{document}|'' command at the conclusion of your source
% document.

% Start the appendix with the ``\verb|appendix|'' command:
% \begin{verbatim}
%   \appendix
% \end{verbatim}
% and note that in the appendix, sections are lettered, not
% numbered. This document has two appendices, demonstrating the section
% and subsection identification method.

% \section{SIGCHI Extended Abstracts}

% The ``\verb|sigchi-a|'' template style (available only in \LaTeX\ and
% not in Word) produces a landscape-orientation formatted article, with
% a wide left margin. Three environments are available for use with the
% ``\verb|sigchi-a|'' template style, and produce formatted output in
% the margin:
% \begin{itemize}
% \item {\verb|sidebar|}:  Place formatted text in the margin.
% \item {\verb|marginfigure|}: Place a figure in the margin.
% \item {\verb|margintable|}: Place a table in the margin.
% \end{itemize}

% \begin{acks}
% \end{acks}

% \clearpage
\bibliographystyle{ACM-Reference-Format}
\bibliography{acmart}

\end{document}